\documentclass[article,onecolumn,superscriptaddress,nofootinbib,floatfix,onecolumn]{revtex4}

\usepackage{amsmath, latexsym, amssymb, hyperref, graphicx, color, multirow, makecell, diagbox}
\usepackage{tabularx}
\usepackage[T1]{fontenc}

\usepackage{dcolumn}
\usepackage{bm}
\usepackage{xspace}

\usepackage{float}
\usepackage{epsfig}
\usepackage{longtable}
\usepackage{rotating}
\usepackage{ulem}
\usepackage{amssymb}
\usepackage{amsmath}
\usepackage{txfonts}
\usepackage{upgreek}
\usepackage{mathtools}
\usepackage{tabularx}
\usepackage{booktabs}

\usepackage{soul} 

\usepackage[capitalize]{cleveref}
\definecolor{nicered}{rgb}{.7,.1,.1}
\definecolor{nicegreen}{rgb}{.1,.5,.1}
\definecolor{darkblue}{rgb}{0,0,.5}
\hypersetup{colorlinks, citecolor=nicegreen,linkcolor=nicered, urlcolor=darkblue}
\usepackage{multirow}
\usepackage{slashed}
\usepackage{verbatim}

\def\mean#1{\ensuremath{\left<#1\right>}}
\def\ttt#1{\texttt{\small #1}}
\def\cO#1{{{\cal{O}}}\left(#1\right)}

\providecommand{\qqbar}{q\overline{q}}
\providecommand{\uubar}{u\overline{u}}
\providecommand{\ddbar}{d\overline{d}}
\providecommand{\ssbar}{s\overline{s}}
\providecommand{\ccbar}{c\overline{c}}

\newcommand{\pp}{p-p}

\providecommand{\pPb}{p-Pb}

\providecommand{\PbPb}{Pb-Pb}

\newcommand{\epem}{e^+e^-}
\newcommand{\ellell}{\ell^+\ell^-}
\newcommand{\mumu}{\mu^+\mu^-}
\newcommand{\tautau}{\tau^+\tau^-}
\newcommand{\tata}{\mathcal{T}}
\newcommand{\ptata}{\mathcal{T}_0}
\newcommand{\otata}{\mathcal{T}_1}

\newcommand{\gaga}{\gamma\gamma}

\newcommand{\jpsi}{\mathrm{J/\psi}}

\newcommand{\etacTwoS}{\mathrm{\eta_{c}(2\mathrm{S})}}

\newcommand{\chicZero}{\mathrm{\chi_{c0}}}
\newcommand{\chicTwo}{\mathrm{\chi_{c2}}}

\newcommand{\llg}{\ensuremath{\epem\to \epem(\gamma),\mumu(\gamma)}\xspace}

\newcommand{\sqrts}{\sqrt{s}}

\newcommand{\sqrtsnn}{\sqrt{s_{_\text{NN}}}}

\newcommand{\pT}{p_\text{T}}
\newcommand{\pTmumu}{p_\mathrm{T}^{\mumu}}

\newcommand{\LumiInt}{\mathcal{L}_{\mathrm{\tiny{int}}}}

\newcommand{\helaconia}{\textsc{HELAC-Onia}}
\newcommand{\madgraph}{\textsc{MadGraph5\_aMC@NLO}}

\newcommand{\gammaUPC}{\ttt{gamma-UPC}}

\def\mean#1{\ensuremath{\left<#1\right>}}

\usepackage{xspace}
\newcommand*{\eg}{e.g.,\@\xspace}
\newcommand*{\ie}{i.e.,\@\xspace}
\newcommand*{\cm}{c.m.\@\xspace}

\begin{document}

\title{Prospects for ditauonium discovery at colliders}

\author{David~d'Enterria}\email{david.d'enterria@cern.ch}
\affiliation{CERN, EP Department, CH-1211 Geneva, Switzerland}
\author{Hua-Sheng~Shao}\email{huasheng.shao@lpthe.jussieu.fr}
\affiliation{Laboratoire de Physique Th\'eorique et Hautes Energies (LPTHE), UMR 7589,\\ Sorbonne Universit\'e et CNRS, 4 place Jussieu, 75252 Paris Cedex 05, France}

\begin{abstract}
\noindent
The feasibility of observing ditauonium, the bound state of two tau leptons, at $\epem$ colliders (BES~III at\,$\sqrts~=~3.78$~GeV, Belle~II at\,$\sqrts~=~10.6$~GeV, a future super tau-charm factory (STCF) at\,$\sqrts \approx 2m_{\tau}$, and the FCC-ee at\,$\sqrts~=~91.2$~GeV) as well as in hadronic and photon-photon collisions at the LHC, is studied. Cross sections and expected yields for spin-0 para- ($\ptata$) and spin-1 ortho- ($\otata$) ditauonium are presented for nine different production and decay processes. Para-ditauonium can be observed at the FCC-ee via photon fusion in its diphoton decay ($\gaga\to\ptata\to\gaga$). Ortho-ditauonium can be observed at STCF via $\epem\to\otata\to\mumu$, where a threshold scan with monochromatized beams can also provide a very precise extraction of the tau lepton mass with at least $\mathcal{O}(25$~keV) uncertainty. 
Observing  pp\,$\to \otata(\mumu)+X$ at the LHC is possible by identifying its displaced vertex with a good control of the combinatorial dimuon background. In addition, we compute the rare decay branching fractions of ditauonium into quarkonium plus~a~photon.
\end{abstract}

\maketitle

\vspace{-0.6cm}

\section{Introduction}

Leptons of opposite electric charge ($\ell^\pm = e^\pm, \mu^\pm, \tau^\pm$) can form short-lived ``onium'' bound states under their quantum electrodynamics (QED) interaction. Among the six possible leptonium states ---$(\epem)$, $(\mu^\pm\mathrm{e}^\mp)$, $(\mumu)$, $(\tau^\pm\mathrm{e}^\mp)$, $(\tau^\pm\mu^\mp)$, and $(\tau^+\tau^-)$--- only the first two, positronium~\cite{Deutsch:1951zza} and muonium~\cite{Hughes:1960zz}, have been observed to date. The lightest system, positronium, has been thoroughly studied for precision QED tests~\cite{Karshenboim:2005iy}, and in searches for violations of the discrete space-time CPT symmetries (where C, P, and T are charge conjugation, parity, and time reversal, respectively) either singly or in various combinations ~\cite{Bernreuther:1988tt,Yamazaki:2009hp}. This work focuses on the heaviest leptonium state, ditauonium, proposed originally in~\cite{Moffat:1975uw,Avilez:1977ai,Avilez:1978sa}, and whose detailed spectroscopic properties~\cite{dEnterria:2022alo} and first feasibility studies for its production via two-photon fusion at $\epem$ and hadron colliders~\cite{dEnterria:2022ysg}, have been recently presented. Its ground state (principal quantum number $n=1$) has two states with total angular momentum $J=0$~and~1, depending on the relative (opposite or parallel) orientation of its constituent tau leptons, known as para- ($\ptata$) and ortho- ($\otata$) ditauonium, respectively. These states, labelled also as $1^1\mathrm{S}_0$ and $1^3\mathrm{S}_1$ using the spectroscopic $n^{2S+1}L_J$ notation (for total spin $S=0,1$ and $L=0,1,...\equiv \mathrm{S},\mathrm{P},...$ orbital angular momentum), have quantum numbers $J^\mathrm{PC}=0^{-+}$ and $1^{--}$, respectively.

The mass of the ditauonium ground state is $m_{_{\tata}} = 2m_\tau + E_\text{bind} = 3553.6962\pm 0.2400$~MeV, with a binding energy of $E_\text{bind} = -\alpha^2 m_\tau/(4n^2) + \mathcal{O}(\alpha^4) = -23.655$~keV at next-to-next-to-leading-order (NNLO) accuracy in the electromagnetic coupling $\alpha= 1/137.036$. This value accounts for a $-115$~eV downwards Lamb shift, and there is a minuscule $\mathcal{O}(3$~eV) hyperfine splitting between the $1^1\mathrm{S}_0$ and $1^3\mathrm{S}_1$ states~\cite{dEnterria:2022alo}. The ditauonium mass uncertainty is fully dominated by the current experimental precision of the tau lepton mass, $m_\tau = 1776.86\pm0.12$~MeV~\cite{Zyla:2020zbs}. The total decay widths of para- and ortho-ditauonium are $\Gamma_\text{tot} = 23.84$ and 31.59~meV, corresponding to lifetimes of $\tauup = \hslash c/\Gamma_\text{tot} = 27.60$ and 20.83~fs, respectively~\cite{dEnterria:2022alo}. Each of the tau leptons forming ditauonium can itself decay weakly with lifetimes $\tauup = 290.3$~fs~\cite{Zyla:2020zbs}, about 10 times longer than that of the combined system, and thus ditauonium can be really produced as a $(\tau^+\tau^-)$ bound state. As a matter of fact, and despite being the shortest-lived leptonium, the ditauonium ground states have lifetimes that are 5 to 8 orders-of-magnitude longer than those of similar charmonium states, with masses also in the 3--4~GeV range, that decay much promptly through the strong interaction. 
The dominant partial decay widths of $\ptata$ into diphotons and $\otata$ into dileptons ($\ell^+\ell^- = \epem$, $\mu^+\mu^-$) are known up to NNLO accuracy~\cite{dEnterria:2022alo}, and amount to
\begin{eqnarray}
\Gamma_{\gaga}(\ptata) &=& \frac{\alpha^5 \, m_\tau}{2\,n^3} \left(1 + \mathcal{O}(\alpha,\alpha^2)\right)= 18.533~\text{meV}\,,\\
\Gamma_{\ell^+\ell^-}(\otata)&=&\frac{\alpha^5\, m_\tau}{6\, n^3}\left(1+\frac{m_\ell^2}{2m_\tau^2}\right)\sqrt{1-\frac{m_\ell^2}{m_\tau^2}}\left(1+\mathcal{O}(\alpha,\alpha^2)\right) = 6.436~\text{meV}\,,
\label{eq:gaga_tata_width}
\end{eqnarray}
corresponding to branching fractions of $\mathcal{B}(\ptata\to\gaga,\otata\to\ell^+\ell^-) = \Gamma_{\gaga,\ell^+\ell^-}/\Gamma_\text{tot} = 77.72\%$ and 20.37\%, respectively, with numerical values given for the $n = 1$ states,\,and theoretical uncertainties commensurate with the last digit quoted. Hadronic decays of the $\otata$ are also possible ($\mathcal{B}(\ptata\to\qqbar)=44.82\%$), but the final states are less clean and subject to larger backgrounds than the leptonic ones, and not considered here.

The motivation for the production and study of ditauonium is threefold. First, ditauonium can provide new high-precision information about fundamental properties of the free tau lepton itself such as its mass, as explained below, and associated precision tests of the SM (\eg\ lepton flavour universality) and of quantities
(\eg\ the CKM $|V_{us}|$ element from $\tau$ decays, 
or empirical relationships claimed among lepton masses~\cite{Koide:1983qe})
that parametrically depend on $m_{\tau}$~\cite{HFLAV:2022pwe}. Second, since the tau lepton is 3500 and 17 times more massive than the electron and muon, respectively, the ditauonium Bohr radius $a_0 = 2/(m_\tau\alpha)= 30.4$~fm is the smallest of all leptonium systems, and its associated minimum ``photon ionization'' energy (Rydberg constant), $R_\infty = \alpha/(4\pi a_0) = 3.76$~keV, is the largest. This implies that $\tata$ is the most strongly bound of all leptonia, and that the investigation of its properties can thereby provide new tests of QED and CPT symmetries at much smaller distances than other exotic atoms. Third, ditauonium features enhanced sensitivity to any physics beyond the standard model (BSM) that is suppressed by powers of $\cO{m_{\ell}/\Lambda_\text{BSM}}$ or affected by uncertainties from hadronic effects, as is the case for, \eg\ positronium or muonic-hydrogen states, respectively. Since various hints of violation of lepton flavour universality have appeared in the last years in studies with free leptons~\cite{HFLAV:2022pwe,Bifani:2018zmi}, it appears relevant to examine if any such effects shows up in the properties of the ditauonium bound states compared to its lighter siblings, positronium and dimuonium.

\begin{figure}[htpb!]
\centering
\includegraphics[width=0.95\textwidth]{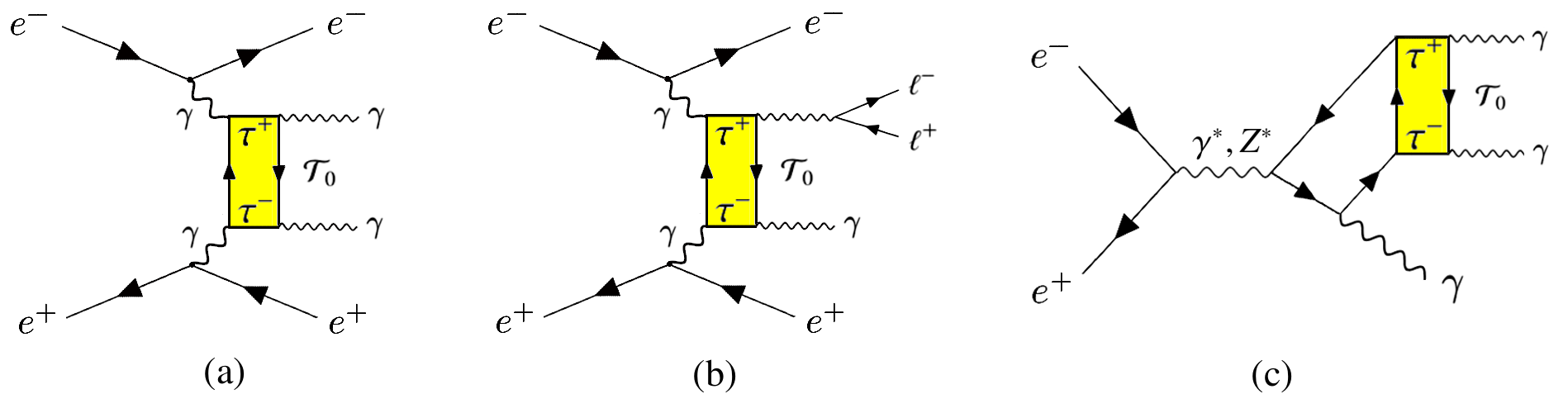}
\includegraphics[width=0.99\textwidth]{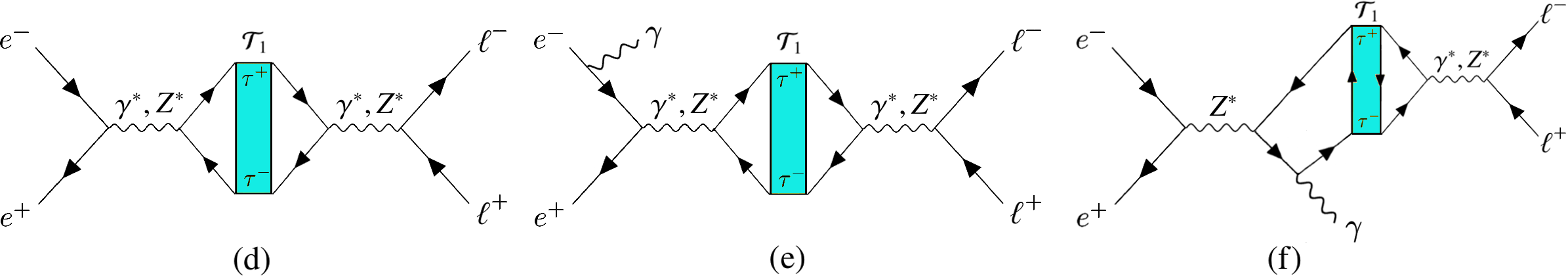}
\caption{Diagrams for para- $\ptata$ (top) and ortho- $\otata$ (bottom) ditauonium production in $\epem$ collisions. Top: Two-photon fusion followed by $\ptata$ (a) diphoton and (b) Dalitz decays, $s$-channel with (c) $\ptata$ diphoton decays with final-state radiation (FSR). Bottom: $s$-channel $\otata$ production with dilepton decays at (d)\,$\sqrts~=~m_{\tata}$, and at\,$\sqrts > m_{\tata}$ with (e) initial-state radiation (ISR) and (f) FSR.
\label{fig:ee_diags}}
\end{figure}

\begin{figure}[htpb!]
\centering
\includegraphics[width=0.99\textwidth]{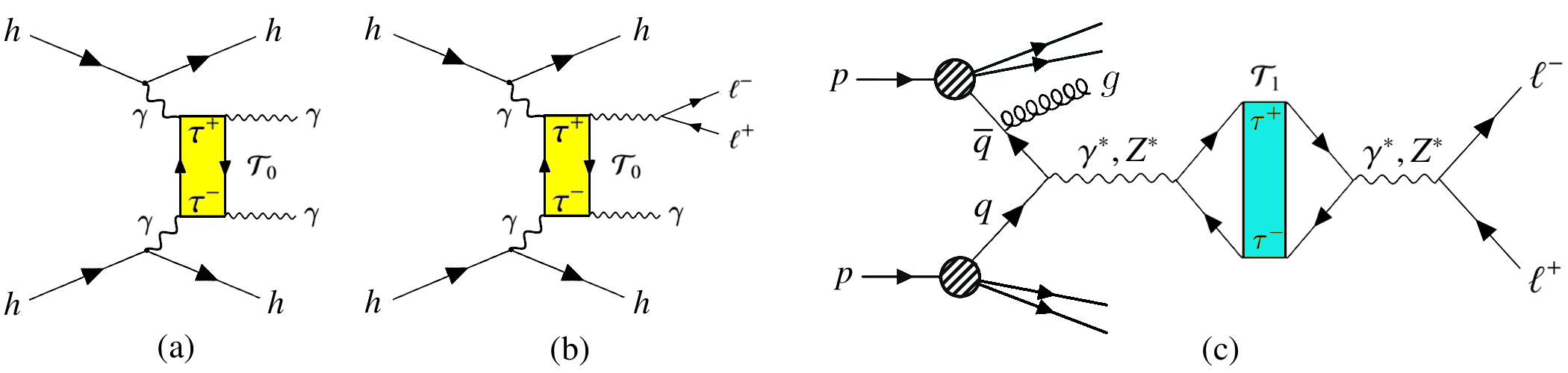}
\caption{Diagrams for ditauonium production at the LHC: exclusive $\ptata$ production via two-photon fusion  (with $\gamma$'s emitted from hadrons $h =$~p,\,Pb) followed by (a) diphoton or (b) Dalitz decays, and (c) Drell--Yan-like $\otata$ production with a recoiling gluon.
\label{fig:pp_diags}}
\end{figure}

So far, the only ditauonium production mode considered in detail in the literature is that of exclusive para-ditauonium in two-photon fusion followed by its diphoton decay $\gaga\to\ptata\to\gaga$~\cite{dEnterria:2022ysg}, shown in the (a) panels of Figs.~\ref{fig:ee_diags} and~\ref{fig:pp_diags}. 
In this work, we consider for the first time seven other ditauonium production and decay modes, including para- and ortho-states (rest of diagrams in both figures). 
Because of C symmetry and the Landau--Yang selection rule~\cite{Landau:1948kw,Yang:1950rg}, production via (quasireal) photon fusion, as well as decay into two photons, is only possible for the spin-0 para-ditauonium state (diagrams (a) and (b) of Figs.~\ref{fig:ee_diags} and~\ref{fig:pp_diags}). Conversely, the $s$-channel production followed by dilepton decays, 
which both proceed through spin-1 virtual photon or Z boson exchanges, are only accessible to ortho-ditauonium (diagrams (d), (e) and (f) of Fig.~\ref{fig:ee_diags}, and (c) of Fig.~\ref{fig:pp_diags}). 
Para-ditauonium can also be produced via $\gamma^*,\,Z^*$ fusion in $\epem$ annihilation provided the final state contains an extra photon (Fig.~\ref{fig:ee_diags} (c)). 

\section{Experimental considerations}

The concrete choices of the ditauonium production and decay processes scrutinized here (Figs.~\ref{fig:ee_diags} and~\ref{fig:pp_diags}) are driven by experimental considerations that can facilitate the observation of the resonance. Potential discovery at four different $\epem$ colliders, as well as at four CERN experiments during the high-luminosity LHC (HL-LHC) phase, will be examined. Among the $\epem$ machines, we consider (i) BES~III running at center-of-mass (\cm) energies\,$\sqrts \approx 3.55,\,3.78$~GeV with total integrated luminosities up to $\LumiInt = 20$~fb$^{-1}$~\cite{BESIII:2020nme}, (ii) a future Super-Tau-Charm-Factory (STCF) at\,$\sqrts \approx m_{\tata}$ integrating up to $\LumiInt = 1$~ab$^{-1}$ of data per year~\cite{Charm-TauFactory:2013cnj,Zhou:2021rgi}, (iii) the Belle~II experiment at Super-KEKB with\,$\sqrts~=~10.6$~GeV and $\LumiInt = 50$~ab$^{-1}$~\cite{Kou:2018nap}, (iv) the future circular collider (FCC-ee) running at the Z pole with\,$\sqrts~=~91.2$~GeV  and $\LumiInt = 50$~ab$^{-1}$~\cite{FCC:2018byv,Abada:2019zxq}. Future linear $\epem$~\cite{ILCInternationalDevelopmentTeam:2022izu,Brunner:2022usy}, as well $\mumu$~\cite{Bartosik:2020xwr}, colliders running at higher energies  will feature integrated luminosities that are likely too low for the production of the rare ditauonium resonances, and will not be considered hereafter.
At the HL-LHC, we consider \pp\ collisions at\,$\sqrts~=~14$~TeV at ATLAS~\cite{ATLAS:2008xda} and CMS~\cite{CMS:2008xjf}, covering central pseudorapidities ($|\eta|\lesssim 3$ for photons and charged leptons) and up to $\LumiInt = 3$~ab$^{-1}$ expected, and the LHCb experiment~\cite{LHCb:2018roe} that covers the forward acceptance ($2 < \eta < 4.5$) with $\LumiInt = 300$~fb$^{-1}$. Production in the LHC fixed-target mode~\cite{Hadjidakis:2018ifr} with the $7$~TeV proton beam on a hydrogen gas target, corresponding to \pp\ collisions at\,$\sqrtsnn=114.6$~GeV with $\LumiInt = 10$~fb$^{-1}$ expected at the LHCb/ALICE interaction points, is also discussed. Photon-fusion processes in ultraperipheral collisions (UPCs)~\cite{Baltz:2007kq} of \pp, proton-lead (\pPb), and lead-lead (\PbPb) at the LHC are also studied. The possibility to reconstruct the very soft decay products (charged leptons and/or photons) of ditauonium produced in UPCs appears only feasible for LHCb in \pp\ collisions, and for the ALICE experiment~\cite{ALICE:2008ngc} in the heavy-ion mode 
with $\LumiInt = 14$~nb$^{-1}$ and  0.66~pb$^{-1}$ for \PbPb\ and \pPb, respectively~\cite{dEnterria:2022sut}.

In the processes $\gaga\to\ptata$ and $\epem\to\otata$ 
where ditauonium is produced alone at rest, its decay vertex is indistinguishable from the primary collision vertex. This holds even in the presence of longitudinal boosts because displaced $\tata$ decays along the beam axis fall within the typical collision spot sizes in the $z$ direction (\eg\ at Belle~II and FCC-ee, the luminous region has a typical $\mathcal{O}(350~\mu$m) $z$-width). For these cases, the ditauonium decay products are swamped by much larger prompt diphoton and/or dilepton (plus photon) backgrounds, and its observation requires the largest possible integrated luminosities. The measurement through resonant $s$-channel production, $\epem\to\otata\to\ell^+\ell^-$ (Fig.~\ref{fig:ee_diags} (d)), relies on counting the number of dilepton events and observing an excess from the ortho-ditauonium decay above the expectations from background $\epem\to\ell^+\ell^-$ processes that have orders-of-magnitude larger cross sections. Maximizing the signal counts requires also to run $\epem$ collisions with \cm\ energies as close as possible to the pole mass of the very narrow $\tata$ Breit--Wigner (BW) peak. 

For the rest of the processes, the presence of decay charged leptons accompanied by the additional production of a photon or a jet (in the \pp\ case), may facilitate the measurement of ditauonium by exploiting its finite decay length (leading to an exponential tail of secondary vertices with proper $\mean{L} = c\tau = 6.2,\,8.2~\mu$m decay lengths for $\otata$ and $\ptata$, respectively), and the boost given by the extra photon $\gamma$ or jet $j$ emitted. Both considerations lead to a potentially visible decay away from the $\epem$ or \pp\ interaction point at an average transverse distance of $\mean{L_{xy}} = \text{boost}_{\perp}\times c\tau$, free from very large prompt SM physics backgrounds, such as \llg processes.
The experimental signature is similar to that of low-mass long-lived dark photon searches via $A'\to\ell^+\ell^-$ decays at $\epem$ and \pp\ colliders~\cite{BaBar:2014zli,BESIII:2017fwv,Kou:2018nap,LHCb:2019vmc}, with the advantage that the mass, width, and lifetime of the $\tata$ particle are known. Transverse ditauonium boosts of the order of 
$(\beta\gamma)_{\perp} = p_{\mathrm{T},\tata}/m_{\tata} \gtrsim 2$ are needed in order to identify the tail of secondary dilepton-decays vertices in the standard silicon trackers with typical vertex position resolutions of the order of $\delta L_{xy}\approx 10~\mu$m. 
We will therefore provide estimates of the number of ditauonium events with displaced vertices, as a proxy for the signal discovery possibilities.
Although decays with $L_{xy} \gtrsim 30~\mu$m take place 
sufficiently separated from the interaction point to expect negligible backgrounds from prompt SM processes, in particular in $\epem$ collisions, such events will nonetheless be potentially subject to instrumental effects that may difficult its observation. In a real analysis, vertex resolution tails, lepton track misreconstruction, and, in the case of dielectron final states, $\gamma\to\epem$ conversions occurring at larger radii but wrongly extrapolated to a vertex closer to the nominal interaction point, should be taken into account. 
Resolution effects will also slightly reduce the signal efficiency. All such subtle reconstruction effects must be evaluated with full detector simulations, for both signal efficiency and background rejection, that are beyond the scope of this work.

\section{Results}

In the following subsections, we present the expected ditauonium cross sections times branching fractions, and associated number of signal events at various $\epem$ facilities and at the LHC, for all processes of Figs.~\ref{fig:ee_diags} and~\ref{fig:pp_diags}. The quoted ditauonium yields are for the expected integrated luminosities of each machine including simple, but realistic, detector acceptance and mass resolution effects (both effects only introduce small signal losses for modern detectors, as shown in the more detailed data analyses of the simulated data carried out in~\cite{dEnterria:2022ysg}).
We will focus on the production of the para- and ortho-ditauonium ground states ($n=1$), because excited states (densely spaced a few keV above the $\tata$ mass) have $n^3$ suppressed cross sections~\cite{dEnterria:2022alo}.
For 
various processes, if relevant, we provide realistic estimates of the number of background events ($B$) and of the expected statistical significance of the signal counts ($S$), given by the $S/\sqrt{B}$ ratio in terms of number of standard deviations ($\sigmaup$) above the background-only expectation. For other channels, we provide the expected number of events with a displaced vertex beyond $L_{xy}\approx 30$, and/or 100~$\mu$m, as indicative of the feasibility of its observation.

\subsection{Para-ditauonium via \texorpdfstring{$\gamma\gamma\to\ptata\to\gamma\gamma$}{gammagamma to TT0 to gammagamma}}

The two-photon fusion production of para-ditauonium followed by its diphoton decay, $\gaga\to\ptata\to\gaga$, in $\epem$ and UPCs (panels (a) of Figs.~\ref{fig:ee_diags} and~\ref{fig:pp_diags}) has been studied in detail in~\cite{dEnterria:2022ysg}. Simulated signal and background events have been generated with the \helaconia\ 2.6.6 Monte Carlo (MC) code~\cite{Shao:2012iz,Shao:2015vga}, complemented with the \gammaUPC\ photon fluxes~\cite{Shao:2022cly}. 
Para-ditauonium is implemented as a modified version of the $\eta_c(1S)$ meson at the $\tata$ mass with a width of $2.4\cdot10^{-8}$~MeV (and corresponding lifetime) accounted for by reshuffling the momentum of the resonance according to its associated BW distribution~\cite{Frixione:2019fxg}. The measurement is very challenging as the final state is swamped by pairs of photons from decays of charmonium resonances ($\chicZero, \chicTwo$, and $\etacTwoS$ mesons) plus the light-by-light (LbL) scattering continuum~\cite{dEnterria:2013zqi}. All such background contributions overlap with the $\ptata$ state and have up to 100 times larger expected yields (Table~\ref{tab:xsecs_gaga_T0_gaga}). Evidence ($3\sigmaup$) and discovery ($5\sigmaup$) of para-ditauonium appear only feasible at Belle II and FCC-ee, respectively, by exploiting their full integrated luminosities and carrying out in-situ high-precision measurements of the diphoton widths of the charmonium mesons. Taking advantage of the $\ptata$ displaced decay vertex appears hopeless given its production at rest (mostly only longitudinal boosts are present in the fusion of two quasireal photons) and the coarse vertex pointing capabilities of reconstructed photons.

\setlength{\tabcolsep}{8.5pt}
\begin{table}[htpb!]
\centering
\caption{Photon-fusion production cross sections times diphoton branching fraction, $\sigma\times\mathcal{B}_{\gaga}$, for the para-ditauonium signal and backgrounds (overlapping C-even charmonium resonances, and LbL scattering over $m_{\gaga}\in (m_{_{\tata}} \pm 100$~MeV)) 
at various $\epem$ facilities, and in \pp, \pPb, and \PbPb\ UPCs at the LHC. The last two columns list the total produced $\ptata$ events for the integrated luminosities at each collider, and the expected statistical significance $S/\sqrt{B}$~\cite{dEnterria:2022ysg}.\label{tab:xsecs_gaga_T0_gaga}}
\vspace{0.1cm}
\begin{tabular}{l|ccccc|cc} \hline
Colliding system,\,$\sqrts$, $\LumiInt$, detector & \multicolumn{5}{c|}{$\sigma\times\mathcal{B}_{\gaga}$} & 
$N(\ptata)\times\mathcal{B}_{\gaga}$ & $S/\sqrt{B}$\\
&  $\etacTwoS$ & $\chicZero$ &  $\chicTwo$ & LbL & $\ptata$ & & \\ \hline
$\epem$ at 3.78~GeV, 20 fb$^{-1}$, BES~III & 3.6 ab & 15 ab & 13 ab & 30 ab   &  0.25 ab &  -- & -- \\
$\epem$ at 7~GeV, 1 ab$^{-1}$, STCF & 0.15 fb &  0.23 fb & 0.33 fb & 0.73 fb & 6.0 ab  & 6 & -- \\ 
$\epem$ at 10.6~GeV, 50 ab$^{-1}$, Belle~II & 0.35 fb & 0.52 fb & 0.77 fb & 1.7 fb  & 0.015 fb &  750  & $3\sigmaup$ \\
$\epem$ at 91.2~GeV, 50 ab$^{-1}$, FCC-ee \;& 2.8 fb & 3.9 fb & 6.0 fb & 12 fb & 0.11 fb &  5\,600 & $5\sigmaup$ \\\hline
\pp\ at 14 TeV, 300~fb$^{-1}$, LHCb &  2.0 fb & 2.8 fb & 4.3 fb & 6.3 fb  & 80 ab &  24  & -- \\
\pPb\ at 8.8 TeV, 0.66~pb$^{-1}$, ALICE & 6.3 pb & 8.7 pb & 13 pb & 21 pb  & 0.25 pb &  0.13 & -- \\
\PbPb\ at 5.5 TeV, 14~nb$^{-1}$, ALICE &  15 nb & 21 nb & 31 nb & 62 nb  & 0.59 nb &  6.4 &  -- \\\hline
\end{tabular}
\end{table}

\subsection{Para-ditauonium via \texorpdfstring{$\gamma\gamma\to \ptata \to \ell^+\ell^-\gamma$}{gammagamma to TT0 to e+e-gamma}}
 
The two-fusion production cross sections of $\ptata$ shown in panels (b) of Figs.~\ref{fig:ee_diags} and~\ref{fig:pp_diags}, are computed with the same setup used in the previous section, but now the Dalitz decays are considered with widths~\cite{dEnterria:2022alo}, 
\begin{eqnarray}
\Gamma_{\ell^+\ell^-\gamma}(\ptata)&=&\frac{2\alpha^6 m_\tau}{9\pi n^3}\left[3\ln{\left(\frac{m_\tau}{m_\ell}+\sqrt{\frac{m_\tau^2}{m_\ell^2}-1}\right)}-\left(4-\frac{m_\ell^2}{m_\tau^2}\right)\sqrt{1-\frac{m_\ell^2}{m_\tau^2}}\right]\left(1 + \mathcal{O}(\alpha,\alpha^2)\right)=\left\{\begin{array}{lr} 0.428~\mathrm{meV}, & \ell^\pm = e^\pm\\
0.125~\mathrm{meV}, & \ell^\pm=\mu^\pm \\\end{array}\right.\,,
\label{eq:para_Dalitz}
\end{eqnarray}
where the numerical values are quoted for the ground state ($n = 1$), and lead to branching fractions of $\mathcal{B}(\ptata\to\ell^+\ell^-\gamma) = 1.79\%$ and 0.52\% for $\ell^\pm = e^\pm,\mu^\pm$, respectively. 
Table~\ref{tab:sigma_ptata_Dalitz} lists the expected para-ditauonium production cross sections times Dalitz decay branching fractions, as well as the corresponding signal yields for the colliding systems considered. In UPCs at the LHC, the cross sections are small for the comparatively low $\LumiInt$ values in heavy-ion collisions, and less than one produced event is expected. In $\epem$ collisions, the cross sections are tiny, reaching 3.2~ab at most, but since the integrated luminosities are very large the number of expected events, combining dielectron and dimuon channels, are $\sim$20 and $\sim$160 at Belle~II and FCC-ee, respectively. Unfortunately, the background cross sections from Dalitz decays of charmonium mesons and the $\epem\to\ell^+\ell^-\gamma$ continuum have expected cross sections many orders of magnitude larger than that of the $\ptata\to\ell^+\ell^-\gamma$ signal. Although Dalitz decays of $\etacTwoS$, $\chicZero$, and $\chicTwo$ have not yet been observed, since this decay, $(\ccbar)_{0,2}\to\gamma\gamma^*(\ell^+\ell^-)$, shares the same underlying diagram as the diphoton case, one expects the same relative $S/B$ cross sections shown in Table~\ref{tab:xsecs_gaga_T0_gaga} for the diphoton channel. Given the factor of $\sim$35 lower number of Dalitz, compared to diphoton, signal events, one does not expect any significant $S/\sqrt{B}$. Also, the transverse boost of $\ptata$ produced via photon fusion at Belle~II and FCC-ee is very small, $\beta\gamma\approx 0.06$, and one will not be able to separate its secondary vertex either. Observation of this exotic atom appears therefore unfeasible in this mode.

\setlength{\tabcolsep}{11pt}
\begin{table}[htpb!]
\centering
\caption{Photon-fusion production cross sections times Dalitz decay branching fraction, $\sigma(\ptata)\times\mathcal{B}_{\gamma\ell\ell}$ with $\mathcal{B}_{\gamma\ell\ell}=\mathcal{B}_{\gamma\epem}+\mathcal{B}_{\gamma\mumu}=2.31\%$, and expected number of events for the para-ditauonium signal at various $\epem$ facilities and in ultraperipheral collisions at the LHC.
\label{tab:sigma_ptata_Dalitz}}
\vspace{0.1cm}
\begin{tabular}{l|c|cccc} \hline
Colliding system,\,$\sqrts$, $\LumiInt$, detector & $\sigma(\ptata)\times\mathcal{B}_{\gamma\ell\ell}$ & $N(\ptata)\times\mathcal{B}_{\gamma ee}$ & $N(\ptata)\times\mathcal{B}_{\gamma\mu\mu}$\\\hline
$\epem$ at 3.78~GeV, 20 fb$^{-1}$, BES~III  &  7.2 zb &  -- & -- \\
$\epem$ at 7~GeV, 1 ab$^{-1}$, STCF  &  0.18 ab &  0.14 & 0.04 \\
$\epem$ at 10.6~GeV, 50 ab$^{-1}$, Belle~II & 0.44 ab & 17 & 5\\
$\epem$ at 91.2~GeV, 50 ab$^{-1}$, FCC-ee & 3.22 ab & 125 & 36 \\\hline
\pp\ at 14 TeV, 300~fb$^{-1}$, LHCb  & 2.3 ab & 0.5 & 0.2 \\
\pPb\ at 8.8 TeV, 0.66~pb$^{-1}$, ALICE & 7.5 fb & -- & -- \\
\PbPb\ at 5.5 TeV, 14~nb$^{-1}$, ALICE  & 17.5 pb  & 0.25 & -- \\\hline
\end{tabular}
\end{table}

\subsection{Para-ditauonium via \texorpdfstring{$\epem\to\ptata(\gaga)+\gamma$}{e+e- to TT0(gammagamma)+gamma}, with an FSR photon}

\begin{figure}[htpb!]
\centering
\includegraphics[width=0.6\textwidth]{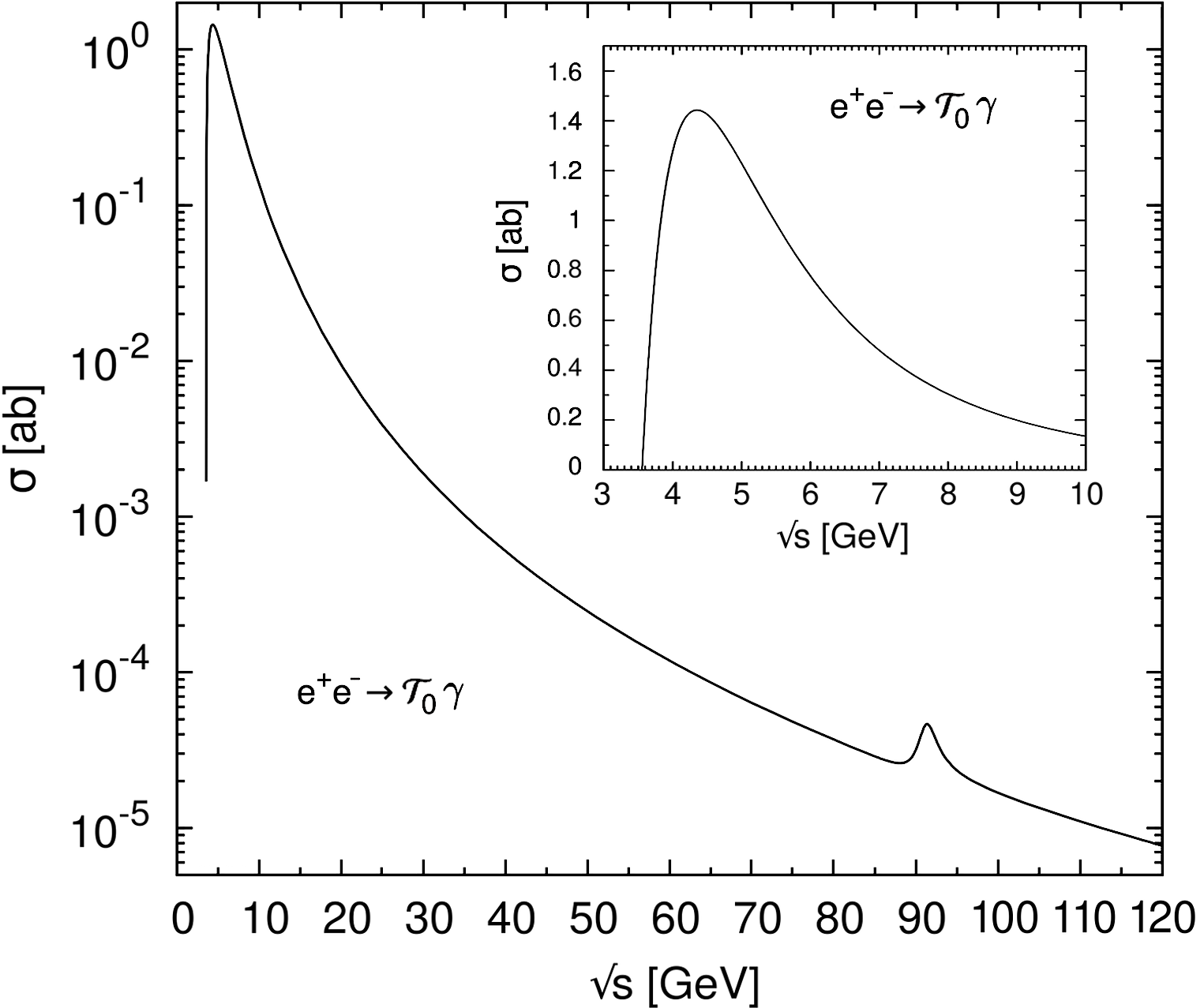}
\caption{Cross section (in ab) for the production of para-ditauonium plus an FSR photon, $\epem\to\ptata+\gamma$, as a function of the $\epem$ \cm\ energy over\,$\sqrts \approx 3$--100~GeV. The inset shows a zoom in the\,$\sqrts = 3$--10~GeV region.
\label{fig:xsee2T0a}}
\end{figure}

The cross section to produce para-ditauonium accompanied with an FSR photon in $\epem$ annihilation via $s$-channel virtual $\gamma$ or Z fusion (Fig.~\ref{fig:ee_diags} (c)) is given by the following expression
\begin{eqnarray}
\sigma(\epem\to \ptata+\gamma) &=&\frac{\pi\alpha^6}{n^3}\frac{m_{\tata}^2}{s^2}\left(1-\frac{m_{\tata}^2}{s}\right)\frac{1}{(s-m_\mathrm{Z}^2)^2+\Gamma_\mathrm{Z}^2 m_\mathrm{Z}^2}\left[\frac{2}{3} m_\mathrm{Z}^4+\frac{8c_w^2-9}{12 s_w^2c_w^2} m_\mathrm{Z}^2 s+\frac{45-84 c_w^2+40 c_w^4}{192 s_w^4 c_w^4}s^2\right],
\label{eq:sigma_T0gamma_FSR}
\end{eqnarray}
where $m_\mathrm{Z}$ and $\Gamma_\mathrm{Z}$ are the Z boson mass and width, and $s_w$ and $c_w$ are sine and cosine of the Weinberg angle. In the infinite $Z$ boson mass limit, the cross section reads
\begin{eqnarray}
\lim_{m_\mathrm{Z}\to \infty}{\sigma(\epem\to \ptata+\gamma)}&=&\frac{2}{3}\frac{\pi\alpha^6}{n^3}\frac{m_{\tata}^2}{s^2}\left(1-\frac{m_{\tata}^2}{s}\right).
\end{eqnarray}
Figure~\ref{fig:xsee2T0a} shows the corresponding cross section as a function of \cm\ energy over\,$\sqrts\approx 3$--100~GeV, with the inset displaying a zoom in the\,$\sqrts = 3$--10~GeV peak region. Starting at the $m_{\tata}$ threshold, the cross section increases with energy up to a maximal value of 1.44 ab reached at\,$\sqrts=\sqrt{3/2\,}m_{\tata}\approx4.3$~GeV, followed by a steady decrease for higher\,$\sqrts$, except for a clear BW structure at\,$\sqrts~=~91.2$~GeV corresponding to the production and ditauonium-mediated decay of the Z boson into three photons (albeit with negligible cross sections)\footnote{The production of para-ditauonium via the radiative decay of bottomonium resonances, or the Higgs boson, is much smaller than the continuum curve shown in Fig.~\ref{fig:xsee2T0a}. The partial widths are given by the expression $\Gamma(\Upsilon(nS)\to\ptata+\gamma) = (2 \alpha^6 m_{\tata}^2 (m_{\Upsilon(nS)}^2-m_{\tata}^2) |R^{\Upsilon(nS)}(0)|^2)/(9m_{\Upsilon(nS)}^6)$, leading to branching fractions of $\Upsilon(1S), \Upsilon(2S), \Upsilon(3S)$ into $\ptata+\gamma$ of $5.4\cdot10^{-12}$, $3.7\cdot10^{-12}$, $4.0\cdot10^{-12}$ respectively, i.e.\ 6 to 7 orders of magnitude smaller than those into $\eta_c+\gamma$ as anticipated from its much larger wave function at the origin~\cite{Eichten:1995ch}. Similarly, the branching fraction of the Higgs boson decay into $\otata+\gamma$ should also be 6 to 7 orders of magnitude smaller than that of the rare $H\to\jpsi+\gamma$ decay.}. In order to retain as much as possible of the cross section, we focus on the dominant $\ptata$ diphoton decay. The corresponding $\epem\to \ptata (\gaga)\gamma$ cross sections and yields for $\epem$ collisions at various colliders are listed in Table~\ref{tab:sigma_ptatagamma}.

\setlength{\tabcolsep}{11pt}
\begin{table}[htpb!]
\centering
\caption{Para-ditauonium production cross sections in $\epem$ annihilation accompanied by an FSR photon, $\epem\to\ptata(\gaga)\gamma$, 
and corresponding expected number of events at various $\epem$ facilities.\label{tab:sigma_ptatagamma}}
\vspace{0.1cm}
\begin{tabular}{l|cc} \hline 
Colliding system,\,$\sqrts$, $\LumiInt$, detector & $\sigma(\ptata+\gamma)\times\mathcal{B}_{\gaga}$ & $N(\ptata(\gaga)+\gamma)$ \\\hline 
$\epem$ at 4.3~GeV, 1 ab$^{-1}$, STCF & 1.1 ab & 1 \\
$\epem$ at 7~GeV, 1 ab$^{-1}$, STCF & 0.37 ab & 0.37 \\
$\epem$ at 3.78~GeV, 20 fb$^{-1}$, BES~III  & 0.69 ab & 0.014 \\
$\epem$ at 10.6~GeV, 50 ab$^{-1}$, Belle~II & 0.085 ab & 4  \\
$\epem$ at 91.2~GeV, 50 ab$^{-1}$, FCC-ee   & $3.6\cdot 10^{-5}$ ab & -- \\
\hline
\end{tabular}
\end{table}

Finite yields are only expected at Belle~II, $N(\ptata(\gaga)+\gamma) = 4$, thanks to its much larger integrated luminosity compared to the rest of low-energy machines. Such low number of signal counts, and a triphoton final-state with poor secondary vertex resolution, precludes a search based on the para-ditauonium displaced vertex tail. 
A search based on identifying an excess above the continuum appears also hopeless as prompt backgrounds sharing the same $3\gamma$ final state 
have notably larger cross sections. Using \helaconia, we find that the production of three photons with at least one pair having an invariant mass within $m_{\tata}\pm 100$~MeV, and the three photons satisfying $10^\circ < \theta^\text{lab}_\gamma < 170^\circ$, yields $\sigma(\epem\to\gamma\gamma\gamma)\times \LumiInt=15~\mathrm{pb}\times 50~\mathrm{ab}^{-1}=7.5\cdot 10^8$ events at Belle II. In addition, the production of a $\chicTwo$ meson plus a photon shares the same final state and mass window, and results in $\sigma(\epem\to \chi_{c2} +\gamma)\times \mathcal{B}_{\chi_{c2}\to \gaga}\times \LumiInt=5090~\mathrm{ab}\times 2.85\cdot 10^{-4}\times 50~\mathrm{ab}^{-1}=70$ counts. With such daunting number of background events, measuring the  $\epem\to \ptata(\gaga)+\gamma$ 
channel considered here appears experimentally unfeasible.
\subsection{Ortho-ditauonium via \texorpdfstring{$\epem\to\otata\to\mumu$}{e+e- to TT1 to mu+mu-}}

\begin{figure}[htpb!]
\centering
\includegraphics[width=0.43\textwidth]{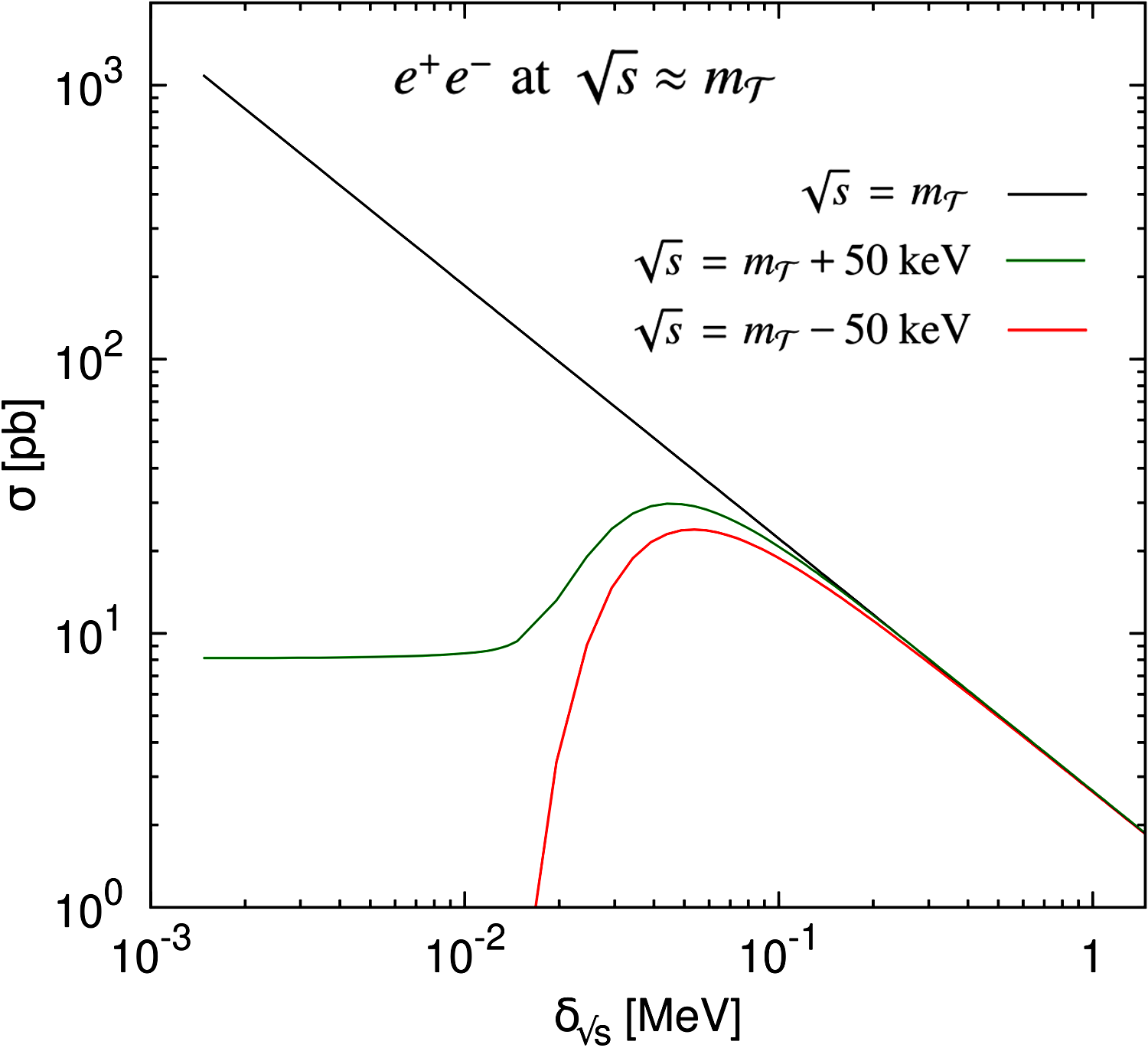}\hspace{1.9mm}
\includegraphics[width=0.55\textwidth]{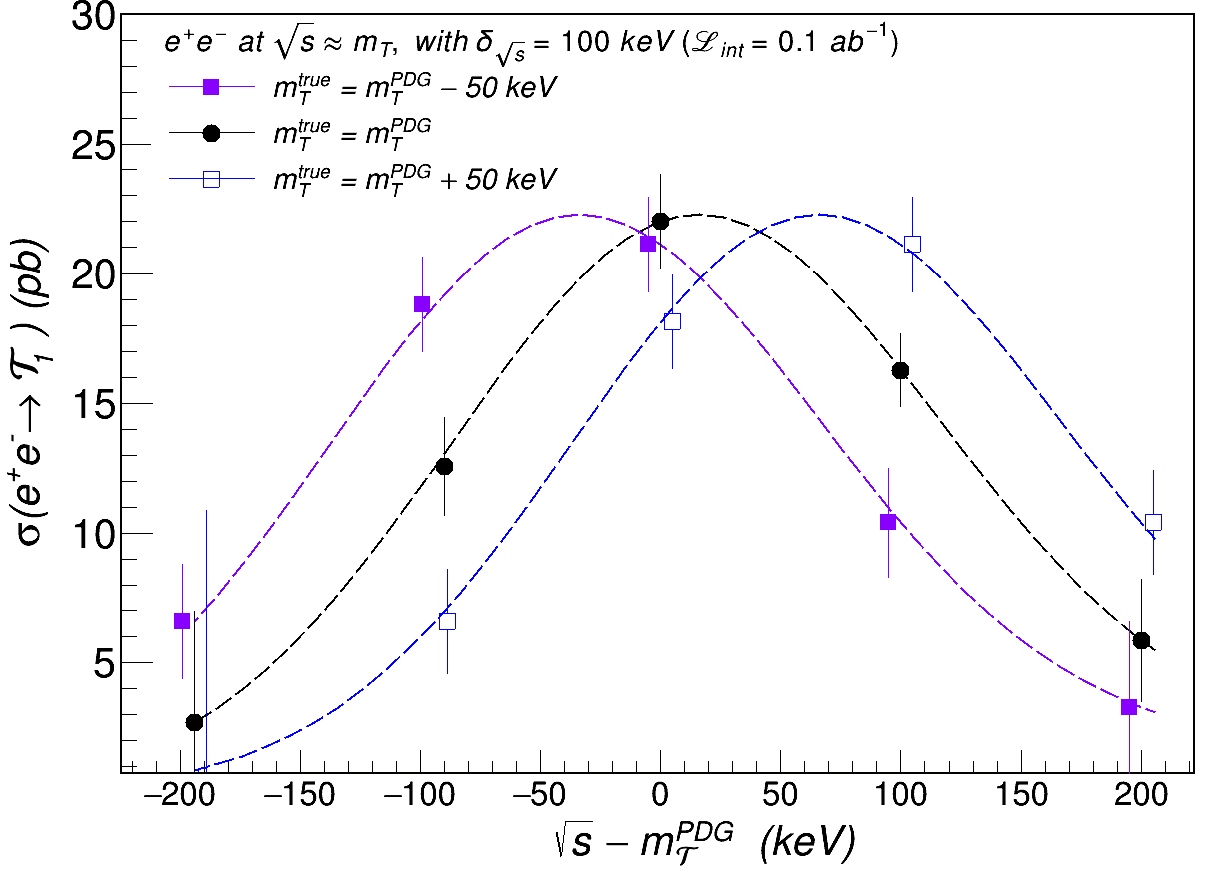}
\caption{Resonant $\sigma(\otata)$ cross section in $\epem$ collisions at\,$\sqrts~=~m_{\tata},m_{\tata}\pm 50~{\rm keV}$ as a function of \cm\ spread $\delta_{\!\sqrts}$ (left), 
and as a function of\,$\sqrts$ minus the $m_{\tata}$ mass for fixed $\delta_{\!\sqrts} = 100$~keV with monochromatized beams (right). In the right plot, the different symbols indicate the expected measured cross sections for 5 $\epem$ scan points (with 0.1~ab$^{-1}$ each) where the actual $\tata$ mass ($m^\text{true}_{\tata}$) coincides with the then-current $m_{\tata}^\text{PDG}$ value (black dots), or is shifted by $\pm 50$~keV (blue and magenta squares) from it  (see text for details).
\label{fig:sigma_TT1_resonant}}
\end{figure}

The  cross section for resonant $\otata$ production in $\epem$ collisions at a \cm\ energy\,$\sqrts$ is theoretically given by the relativistic BW expression:
\begin{eqnarray}
\sigma^\text{ideal}(\epem\to \otata)&=&\frac{12\pi \Gamma_\text{tot}(\otata)\,\Gamma_{\epem}(\otata)}{(s-m_{\tata}^2)^2+\Gamma_\text{tot}^2(\otata)\,m_{\tata}^2}.
\label{eq:sigma_BW}
\end{eqnarray}
For an $\epem$ run exactly at the pole mass,\,$\sqrts=m_{\tata}$, this cross section amounts to $\sigma^\text{ideal}= 12\pi \mathcal{B}_{\epem}(\otata)/m_{\tata}^2 = 236.6~\mu$b. However, similar to the case of resonant Higgs boson production at the FCC-ee, $\epem\to \mathrm{H}$~\cite{dEnterria:2021xij}, two effects reduce substantially this purely theoretical cross section. First, the actual beams are never perfectly monoenergetic but have a distribution of incoming energies that is many orders-of-magnitude wider than the width $\Gamma_\text{tot}(\otata)$ of the very narrow ditauonium resonance. Secondly, the occurrence of ISR photon emission from the $e^\pm$ beams will shift the actual collision energy\,$\sqrts$ below the BW peak. Both effects render the $(s-m_{\tata}^2)^2$ term much larger than $\Gamma_\text{tot}^2(\otata)\,m_{\tata}^2$ in the denominator of Eq.~(\ref{eq:sigma_BW}), and strongly reduce the actual cross section. For a Gaussian $\epem$ \cm\ energy spread of width $\delta_{\!\sqrts}$ (with $\delta_{\!\sqrts}\ll \sqrts$), the actual $s$-channel cross section with both effects convolved can be written as
\begin{eqnarray}
\!\!\!\!\!\sigma^\text{actual}(\epem\to \otata)&=&\frac{12\pi^2\Gamma_{\epem}(\otata)}{m_{\tata}}\int_0^1{\!\!dx_1\!\!\int_0^1{\!\!dx_2 f_{e^-/e^-}(x_1,s)f_{e^+/e^+}(x_2,s)V_2\left(\sqrt{x_1x_2s}; m_{\tata},\Gamma_\text{tot}(\otata),\sqrt{x_1x_2}\delta_{\!\sqrts}\right)}},
\label{eq:xsthr}
\end{eqnarray}
where the $f_{e^-/e^-}$ and $f_{e^+/e^+}$ structure functions describe the distribution of energy of each $e^\pm$ beam reduced by a fraction $x_{1,2}$ after collinear initial photon emission, and 
where the relativistic Voigtian function $V_2$ reads
\begin{eqnarray}
V_2(E;m,\Gamma,\delta)&=&\int_{-\infty}^{+\infty}{d\tilde{E}\frac{m\Gamma}{\pi}\frac{1}{(\tilde{E}^2-m^2)^2+\Gamma^2m^2}\frac{1}{\sqrt{2\pi}\delta}e^{-\frac{(\tilde{E}-E)^2}{2\delta^2}}},
\label{eq:Voigtdist}
\end{eqnarray}
which can be expressed in terms of the complementary error function~\cite{Kycia:2017gjn}. In order to carry out the two integrations in Eq.~(\ref{eq:xsthr}), we use the leading-logarithmic expressions for $f_{e^\pm/e^\pm}$, and the parametrization of the fractions of beam energies $x_{1,2}$, of Ref.~\cite{Frixione:2021zdp}. With those ingredients at hand, one can compute the $\epem\to \otata$ cross sections for any\,$\sqrts$ and $\delta_{\!\sqrts}$ values. Figure~\ref{fig:sigma_TT1_resonant} (left) shows the resonant $\otata$ cross section in $\epem$ collisions at\,$\sqrts = m_{\tata}$ and $m_{\tata}\pm 50~{\rm keV}$ as a function of the \cm\ spread $\delta_{\!\sqrts}$. Assuming that one runs exactly at the $m_{\tata}$ mass, the cross section decreases very rapidly, following a $1/\delta_{\!\sqrts}$ dependence (top black curve). If the \cm\ energy of the $\epem$ collision misses the mass of the resonance by $+50$~keV (middle green curve) or by $-50$~keV (bottom red curve), the cross section peaks at $\sigma^\text{actual}(\epem\to \otata)\approx 20$~pb for a $\delta_{\!\sqrts}\approx 50$~keV spread, and decreases to the left and right of it. The drop in the cross section in the region $\delta_{\!\sqrts}\lesssim 50$~keV is much smaller for the $+50$-keV shift, than for the $-50$-keV case, because running above the resonance profits from possible ISR radiative returns that bring the $\epem$ \cm\ energy onto the BW peak.

The experimental observation of resonant $\otata$ production is a ``counting experiment'' whereby one measures the number of dilepton events searching for an excess issuing from the ortho-ditauonium decays above the expectations from  the background-only $\epem\to\ell^+\ell^-$ processes. We focus on the $\otata\to\mumu$ decay mode which has smaller continuum backgrounds than the $\otata\to\epem$ channel. As discussed in the direct Higgs boson production in $\epem$ collisions at\,$\sqrts = m_\mathrm{H}$ at the FCC-ee~\cite{dEnterria:2021ljz}, in order to maximize the cross section one needs to: (i) run at a\,$\sqrts$ value as close as possible to the true ditauonium mass $m_{\tata}$ (known today with a 250-keV uncertainty, from twice the tau-lepton mass value precision), (ii) to be able to measure and calibrate the actual machine\,$\sqrts$ within the same high accuracy, and (iii) to reduce the \cm\ energy spread $\delta_{\!\sqrts}$ down to values of the same order. Regarding the first point, the BES~III collaboration performed a fine mass scan experiment in 2018 with five points around the tau pair production threshold with a\,$\sqrts$ spread of $\delta_{\!\sqrts} = 1.24$~MeV, and a total integrated luminosity of 140~pb$^{-1}$~\cite{Achasov:2019rdp}. The final expected uncertainty in $m_{\tau}$ will be around 50~keV. Similarly, Belle~II forecasts to reach tau mass uncertainties below 150~keV (systematic and statistical sources combined) with about 300~fb$^{-1}$ of data~\cite{Belle-II:2020wbx}. Combining these results, plus all others from the PDG world average~\cite{Zyla:2020zbs}, it is not inconceivable to reach a precision of $\mathcal{O}(50~\text{keV})$ in the tau mass (\ie\ twice this value, $\mathcal{O}(100~\text{keV})$, in the $\tata$ mass) whenever a dedicated search for the ditauonium is carried out at a future STCF. 
Regarding the\,$\sqrts$ calibration, the BES-III experiment has already achieved a high accuracy, at the level of $2\cdot10^{-5}$, by exploiting the measurement of monochromatic laser photons backscattered from the $e^\pm$ beams in its beam energy measurement system (BEMS)~\cite{Abakumova:2011rp}. The application of the same technique at three different lepton colliders can reach accuracies of the actual \cm\ energy not worse than 50~keV~\cite{Achasov:2020iou}. Last but not least, on the \cm\ energy spread, the possibility of monochromatization of electron/positron beams has been considered several times in the literature~\cite{Kirkby:1996qt,Bogomyagkov:2017uul,Telnov:2020rxp}, with values of $\delta_{\!\sqrts} = 50$~keV at the tau-pair production threshold theoretically achievable.

\setlength{\tabcolsep}{5pt}
\begin{table}[htpb!]
\centering
\caption{Cross sections and expected number of events for the $s$-channel production of ortho-ditauonium ($\otata$), and for the $\tautau$ and (background) $\mumu$ continua, in $\epem$ at\,$\sqrts\approx m_{\tata}$ at various facilities. The last column lists the expected signal statistical significance.
\label{tab:xsecs_scan}}
\vspace{0.1cm}
\begin{tabular}{l|ccc|ccc|c} \hline
Colliding system,\,$\sqrts$ ($\delta_{\!\sqrts}$ spread), $\LumiInt$, experiment & \multicolumn{3}{c|}{$\sigma$} & \multicolumn{3}{c|}{$N$}  & $S/\sqrt{B}$ \\
& $\otata$ &  $\tautau$ & $\mumu$ & $\otata$ & $\otata\to\mumu$ & $\mumu$  &  \\ \hline
$\epem$ at 3.5538~GeV (1.47~MeV), 5.57 pb$^{-1}$, BES~III & 1.9 pb & 117 pb & 6.88 nb &  10.4 & 2.1 & 38\,300 & $0.01\sigmaup$\\
$\epem$ at\,$\sqrts \approx m_{\tata}$ (1.24~MeV), 140 pb$^{-1}$, BES~III & 2.2 pb &  103 pb & 6.88 nb &  310 & 63 & $9.63\cdot 10^5$ & $0.06\sigmaup$\\
$\epem$ at\,$\sqrts \approx m_{\tata}$ (1~MeV), 1 ab$^{-1}$, STCF & 2.6 pb & 95 pb & 6.88 nb & $2.6\cdot 10^6$ & $5.3\cdot 10^5$ & $6.88\cdot 10^9$  & $6.4\sigmaup$ \\
$\epem$ at\,$\sqrts \approx m_{\tata}$ (100~keV), 0.1 ab$^{-1}$, STCF & 22 pb & 46 pb & 6.88 nb &  $2.2\cdot 10^6$ & $4.5\cdot 10^5$ & $6.88\cdot 10^8$  & $17\sigmaup$ \\
\hline
\end{tabular}
\end{table}

Table~\ref{tab:xsecs_scan} lists the expected resonant $\otata$ cross sections and number of events at various $\epem$ facilities. We list first the two BES-III $m_{\tau}$ scan runs performed around the $\tau$-pair threshold in 2011 and 2018 with $\LumiInt = 5.57, 140$~pb$^{-1}$ and $\delta_{\!\sqrts}=1.469, 1.24$~MeV, respectively~\cite{BESIII:2014srs,Achasov:2019rdp}. 
For each of these two cases, we expect about 2 and 60 $\epem\to\otata(\mumu)$ events produced respectively, which are too small numbers to be observed on top of the orders-of-magnitude larger dimuon continuum background. On the other hand, the STCF is expected to integrate 1~ab$^{-1}$ around the $\epem\to\tautau$ threshold with a default $\delta_{\!\sqrts} \approx 1~\mathrm{MeV}$ spread~\cite{Charm-TauFactory:2013cnj,Zhou:2021rgi}. One single run under these conditions will produce more than half-a-million $\otata$ particles decaying into dimuons, enough to observe its production with a significance around $S/\sqrt{B} = 6.5\sigmaup$. For the STCF facility, we consider in addition the possibility to monochromatize the beams down to $\delta_{\!\sqrts}\approx 100~\mathrm{keV}$, albeit with a factor of ten loss in the integrated luminosity. Such assumptions should be considered as conservative given that spreads as low as $\delta_{\!\sqrts}\approx 50$~keV are expected at these colliding energy (albeit with corresponding losses in the beam luminosities)~\cite{Bogomyagkov:2017uul,Telnov:2020rxp}. With such a monochromatization working point, we expect more than 2.2 million ditauonium events produced per 0.1-ab$^{-1}$ scan point that would allow a detailed study of the resonance. In particular, if the \cm\ energy is calibrated to within 50-keV with the BEMS (or any similar) technique, one can then determine the ditauonium mass within this precision by carrying out a scan with \eg\ five collision points around the production threshold even with a reduced 0.1~ab$^{-1}$ integrated luminosity per point. With such a setup, one would be able to determine the position of the $\otata$ mass peak within an accuracy just driven by the $\mathcal{O}(50$~keV) calibration value. This is illustrated in Fig.~\ref{fig:sigma_TT1_resonant} (right) where the $s$-channel cross section around the $m_{\tata}$ pole for a fixed $\delta_{\!\sqrts} = 100$~keV spread is shown. The three sets of 5-scan points show the expected extracted $\otata$ cross sections (with error bars approximating the statistical uncertainties only) for three possible cases where the actual ditauonium mass corresponds to the assumed PDG mass at the moment of performing the experiment, or where it is actually shifted by $\pm50$~keV with respect to it. The plot shows clearly that a Gaussian fit (dashed curves) to whichever measured set of five cross section points should provide a peak value whose position will only be driven by the accuracy in the knowledge of the \cm\ energy calibration. Such an extraction of the $m_{\tata}$ value with $\pm50$~keV uncertainty would allow a direct determination of the $\tau$ lepton mass, $m_{\tau} = \left(m_{\tata}-E_\text{bind}\right)/2$ (with $E_\text{bind} = -23.655$~keV), to within $\pm25$~keV. Given that the tau lepton decays into final states with invisible neutrinos that complicate the accurate measurement of its mass (and that any approach based on a high-luminosity $\tautau$ threshold scan will inevitably produce ditauonium, even without monochromatized beams, as shown in Table~\ref{tab:xsecs_scan}), no other $m_{\tau}$-determination method can likely match the precision and accuracy of the ditauonium-mass presented here. We note that the $\otata$ cross section is about 2\% (50\% for monochromatized beams) of the total tau-pair production for the $\epem$ collisions at $\sqrts \approx 2m_{\tau}$ considered in Table~\ref{tab:xsecs_scan}. Namely, about 2\% of the $\tau$ pairs produced in standard $\epem$ threshold scans will form an ortho bound state.
It is obvious from this study that ditauonium should be an integral part of the experimental programme of precision studies of the tau lepton at any future STCF.

\subsection{Ortho-ditauonium via \texorpdfstring{$\epem\to \otata(\ellell)$}{e+e- to TT1(l+l-)} in ISR collisions}

The cross section to produce ortho-ditauonium in $\epem$ annihilation at\,$\sqrts > m_{\tata}$ after $\gamma$ ISR emission (Fig.~\ref{fig:ee_diags} (e), with a collinear photon) is given by
\begin{eqnarray}
\sigma(\epem\to\otata,\,\text{ISR}) &=& \frac{\pi^2\alpha^5}{n^3}\frac{1}{s}\frac{1}{(m_{\tata}^2-m_\mathrm{Z}^2)^2}\left[m_\mathrm{Z}^4+\frac{8c_w^2-9}{8 s_w^2 c_w^2}m_\mathrm{Z}^2m_{\tata}^2+\frac{45-84 c_w^2+40 c_w^4}{128 s_w^4 c_w^4}m_{\tata}^4\right]\,\mathcal{L}_{\epem}\left(\frac{m_{\tata}^2}{s},s\right),
\end{eqnarray}
with the same parameters defined in Eq.~(\ref{eq:sigma_T0gamma_FSR}), and where the last factor encodes the change in the ditauonium cross section due to the emission of ISR photon(s). This latter factor is defined as an effective $\epem$ luminosity in terms of the electron and positron structure functions, $f_{e^-/e^-}(x,s)$ and $f_{e^+/e^+}(x,s)$,
\begin{eqnarray}
\mathcal{L}_{\epem}(z,s)&=&\int_{z}^{1}{\frac{dx}{x} f_{e^-/e^-}\left(x,s\right)f_{e^+/e^+}\left(\frac{z}{x},s\right)},
\end{eqnarray}
where $x$ is the fraction of the beam momentum carried out by the $e^\pm$ after collinear ISR emission.
This effective luminosity can be evaluated either via MC methods~\cite{Shao:2014rwa} or in a parton-distribution-function (PDF)-like approach~\cite{Kuraev:1985wb,Beenakker:1996kt,Frixione:2019lga,Bertone:2019hks}.
Again, in the infinite Z boson mass limit, a more handy formula is available
\begin{eqnarray}
\lim_{m_\mathrm{Z}\to \infty}{\sigma (\epem\to\otata,\,\text{ISR})}&=& \frac{\pi^2 \alpha^5}{n^3 s}\,\mathcal{L}_{\epem}\left(\frac{m_{\tata}^2}{s},s\right).
\end{eqnarray}
Here, we use the $\mathcal{L}_{\epem}(z,s)$ effective luminosity at leading-logarithmic accuracy from Eq.~(67) of Ref.~\cite{Beenakker:1996kt}, with the so-called `mixed' prescription that partially accounts for a few extra non-leading terms in the cross section. 
Table~\ref{tab:xsecs_ISR} lists the corresponding ISR ortho-ditauonium cross sections ($n=1$) at four different $\epem$ colliders. The largest signal cross sections, in the few~fb range, are for BES-III and STCF running at a\,$\sqrts$ above, but not too far from, the $\tau^+\tau^-$ production threshold. About 25 and 1200 $\otata$ events are expected in the combined dilepton decays at each machine, respectively. At Belle~II, despite two orders-of-magnitude smaller cross sections, one expects about 750 dilepton signal events thanks to its much larger integrated luminosities, whereas the FCC-ee energies are too large for ISR ditauonium production. In all cases, the number of prompt dilepton continuum events below the $\otata$ peak is, however, orders-of-magnitude larger, and the possibility to observe its displaced vertex decay is negligible given that the ISR photons are emitted collinear to the beam and do not impart any appreciable transverse boost to the resonance. In the next section, the more favorable case where the ISR/FSR $\gamma$ is emitted at a large angle is discussed.

\setlength{\tabcolsep}{7.2pt}
\begin{table}[htpb!]
\centering
\caption{Production cross sections and expected yields for ortho-ditauonium and prompt dilepton backgrounds in $\epem$ annihilation with ISR at four different colliders. 
The $\ellell$ backgrounds are computed within $m_{\ellell}\in (m_{\tata}\pm5~\mathrm{MeV})$ and $10^\circ<\theta^\text{lab}_{\ell^\pm}<170^\circ$.
\label{tab:xsecs_ISR}}
\vspace{0.1cm}
\begin{tabular}{l|ccc|cccc} \hline
Colliding system,\,$\sqrts$, $\LumiInt$, detector & \multicolumn{3}{c|}{$\sigma$} & 
\multicolumn{4}{c}{$N$}\\
 & $\otata$ & $\mumu$ & $\epem$ & $\otata\to\mumu$ & $\otata\to\epem$ & $\mumu$ & $\epem$\\ \hline
$\epem$ at 3.78~GeV, 20 fb$^{-1}$, BES~III & 3.0 fb & 18.3 pb & 6.4 nb & 12 & 12 & $3.7\cdot 10^5$ & $1.3\cdot 10^8$\\
$\epem$ at 3.78~GeV, 1 ab$^{-1}$, STCF & 3.0 fb & 18.3 pb & 6.4 nb & 610 & 610 & $1.8\cdot 10^7$ & $6.4\cdot 10^{9}$\\
$\epem$ at 10.6~GeV, 50 ab$^{-1}$, Belle~II & 36 ab & 223 fb & 10.5 pb & 370 & 370 & $1.1\cdot 10^{7}$ & $5.3\cdot 10^{8}$\\
$\epem$ at 91.2~GeV, 50 ab$^{-1}$, FCC-ee \;& 0.61 ab & 3.8 fb & 10 fb & 6 & 6 & $1.9\cdot 10^5$ & $5.0\cdot 10^5$\\
\hline
\end{tabular}
\end{table}

\subsection{Ortho-ditauonium via \texorpdfstring{$\epem\to\otata(\ellell)+\gamma$}{e+e- to TT1(l+l-)+gamma} with a large-angle photon}

In order to transversely boost the ortho-ditauonium and facilitate the observation of its displaced decay vertex, we consider $\epem\to \otata+\gamma$ where the emitted photon has a large scattering angle (\ie\ it is not collinear with the beams, $\theta\neq0,\pi$) in the rest frame of the $\otata$ and $\gamma$ system (Figs.~\ref{fig:ee_diags} (e) and (f) for a non-collinear photon). In the above frame, the differential cross section with respect to the cosine of the relative angle $\theta$ is
\begin{eqnarray}
\frac{d\sigma(\epem\to \otata+\gamma)}{d\cos{\theta}}&=&\frac{\pi\alpha^6}{512 n^3 s_w^4 c_w^4}\frac{x_s}{s}\left\{\frac{s^2}{(s-m_\mathrm{Z}^2)^2+\Gamma_\mathrm{Z}^2m_\mathrm{Z}^2}\left[\left(1-4s_w^2+8s_w^4\right)\left(1-x_s\right)\left(1+2x_s+(1-2x_s)\cos^2{\theta}\right)\right.\right.\nonumber\\
&&\left.-\frac{4\left(1+x_s\right)\left(x_s-x_z\right)\left((1+8s_w^4)x_z-8s_w^2c_w^2\right)}{x_z\left(1-x_z\right)}\right]\\
&&\left.+\frac{2\left[(1+x_s)^2+(1-x_s)^2\cos^2{\theta}\right]\left[128s_w^4c_w^4-16s_w^2c_w^2(1+8s_w^2)x_z+(1+4s_w^2+40s_w^4)x_z^2\right]}{\left(1-\cos^2{\theta}\right)x_s\left(1-x_s\right)\left(1-x_z\right)^2}\right\},\nonumber\label{eq:xsorthoA}
\end{eqnarray}
where $x_s=\tfrac{m_{\tata}^2}{s}$ and $x_z=\tfrac{m_{\tata}^2}{m_\mathrm{Z}^2}$. For $m_\mathrm{Z}\to \infty$ or $x_z\to 0$, the cross section has the asymptotic limit
\begin{eqnarray}
\lim_{m_\mathrm{Z}\to \infty}{\frac{d\sigma(\epem\to \otata+\gamma)}{d\cos{\theta}}}&=&\frac{\pi\alpha^6}{n^3}\frac{1}{s}\frac{(1+x_s)^2+(1-x_s)^2\cos^2{\theta}}{2(1-\cos^2{\theta})(1-x_s)}.
\end{eqnarray}
The $\otata$ scalar three-momentum in the same frame is fixed at $|p| = \tfrac{s-m_{\tata}^2}{2\sqrts}$, where $s$ is the invariant mass square of the final $\otata+\gamma$ system. At Belle II collisions with\,$\sqrts=10.6$~GeV, the $\otata$ three-momentum in the rest frame is $|p|\approx 4.7$~GeV. One can integrate Eq.~(\ref{eq:xsorthoA}) over $\cos{\theta}$ from $-1+\varepsilon_1$ to $1-\varepsilon_2$, where $\varepsilon_{1,2}$ are positive real numbers smaller than one, to avoid the collinear divergence of the $\frac{1}{1-\cos^2{\theta}}$ term for $\cos{\theta}\to \pm 1$. The derived cross sections and yields values for $\epem\to \otata+\gamma$ expected at different facilities can only be a fraction of the ISR ones quoted in Table~\ref{tab:xsecs_ISR}, because the underlying physical process is the same except that now we require a hard photon within the detector acceptance that boosts the $\otata$ in the opposite direction. Given the low number of events already expected for ISR collisions at BES-III and FCC-ee (Table~\ref{tab:xsecs_ISR}), there is no chance to see any $\otata+\gamma$ event of this sort at both machines. Similarly, the STCF collision energies or integrated luminosities will not be adequate to produce a photon hard enough to boost the ditauonium resonance. One is therefore constrained to just consider the Belle~II case, for which the $\otata+\gamma$ cross sections, where the photon is emitted within the detector acceptance ($10^\circ<\theta^\text{lab}_\gamma<170^\circ$) are given in Table~\ref{tab:xsecs_FSR}. The prompt dilepton-plus-photon backgrounds are many orders of magnitude larger so only displaced vertices will make the $\otata$ visible. Simulated $\epem\to\otata(\ellell)+\gamma$ signal events are generated with \helaconia\ and passed through the Belle-II acceptance for the final-state particles. Combining dielectron and dimuon decays, the total expected number of $\otata(\ellell)$ events is $\sim$150 with only one event expected in the tail of the displaced $L_{xy}$ decays. The measurement at Belle-II is therefore certainly challenging, but given that the mass and width of the resonance are known, the search at $L_{xy}$ values away from the interaction point would be worth a try.
 
\setlength{\tabcolsep}{10pt}
\begin{table}[htpb!]
\centering
\caption{Production cross sections and expected yields for ortho-ditauonium and prompt dilepton backgrounds ($10^\circ<\theta^\text{lab}_{\ell^\pm}<170^\circ$), plus a hard large-angle photon ($10^\circ<\theta^\text{lab}_\gamma<170^\circ$), in $\epem$ annihilation at Belle~II. 
The prompt $\ellell\gamma$ backgrounds are computed for $m_{\ellell}\in (m_{\tata}\pm5~\mathrm{MeV})$ (for $\epem\gamma$, we also require the angle between $\gamma$ and $e^\pm$ to be $10^\circ<\Delta\theta^{\text{lab}}_{\gamma e^\pm}<170^\circ$ to avoid final-state collinear divergences). The last columns list the expected $\otata(\ellell)+\gamma$ events with a displaced vertex beyond 30~$\mu$m.
\label{tab:xsecs_FSR}}
\vspace{0.1cm}
\begin{tabular}{l|ccc|cccc} \hline
Colliding system,\,$\sqrts$, $\LumiInt$, detector & \multicolumn{3}{c|}{$\sigma$} & 
\multicolumn{2}{c}{$N(\otata+\gamma)$} & \multicolumn{2}{c}{with $L_{xy}> 30~\mu$m}\\
 & $\otata+\gamma$ & $\mumu+\gamma$ & $\epem+\gamma$ & $\mumu$ & $\epem$ & $\mumu$ & $\epem$\\ \hline
$\epem$ at 10.6~GeV, 50 ab$^{-1}$, Belle~II & 7.1 ab & 51 fb & 1.9 pb & 73 & 73 & 0.5 & 0.5 \\
\hline
\end{tabular}
\end{table}


\subsection{Ortho-ditauonium via \texorpdfstring{pp\,$\to \otata(\mumu)+X$}{pp to TT1(mu+mu-)}}

Ortho-ditauonium can be produced in \pp\ collisions 
through the Drell--Yan-like $q\bar{q}\to \gamma^*\to \otata(\mumu)+X$ process shown in Fig.~\ref{fig:pp_diags} (c). At LO in pQCD, the cross section for p\,p\,$\to \otata+X$ at a given nucleon-nucleon \cm\ energy\,$\sqrtsnn$ can be obtained from
\begin{eqnarray}
\sigma^\mathrm{LO}&=&\sum_{q=u,d,s}{\int_z^1{dx_1\int_{z/x_1}^{1}{dx_2}{\left[f_{q/p}(x_1,\mu_F)f_{\bar{q}/p}(x_2,\mu_F)+f_{\bar{q}/p}(x_1,\mu_F)f_{q/p}(x_2,\mu_F)\right]\delta\left(1-\tfrac{m_{\tata}^2}{\hat{s}}\right)\hat{\sigma}_{\qqbar}(\hat{s},m_{\tata})}}},
\end{eqnarray}
where $f_{q/p}(x,\mu_F^2)$ are PDFs evaluated at parton momentum fractions $x_i$ and factorization scale $\mu_F$, $\hat{s}=x_1x_2 s_\mathrm{NN}$ is the squared partonic invariant mass, $z=\tfrac{m_{\tata}^2}{s_\mathrm{NN}}$, and the partonic cross section of quarks of charge $Q_q$ is
\begin{eqnarray}
\hat{\sigma}_{\qqbar}(\hat{s},m_{\tata})&=&\frac{\pi^2\alpha^5}{3n^3m_{\tata}^2}Q_q^2.
\end{eqnarray}
The NLO cross sections for inclusive $\otata$ production at the LHC are obtained using the expressions above with the {\tt CT14nlo} PDFs~\cite{Dulat:2015mca} and an NLO/LO $K\approx 1.4$ ratio computed with \madgraph~v2.6.6~\cite{Alwall:2014hca} from the Drell--Yan p\,p\,$\to \gamma^* (j)\to \mumu (j)$ processes with the invariant mass of the dimuon system set at $m_{\tata}$, and the central $\mu_F$ scale set at the transverse mass of the pair.
The corresponding ortho-ditauonium ($n=1$) cross sections are listed in the first column of Table~\ref{tab:xsec_pp} for \pp\ collisions at\,$\sqrts=14$~TeV, and\,$\sqrtsnn=114.6$~GeV (fixed target). The quoted $\sigma_\mathrm{NLO}(\otata+X)$ uncertainties of 30--40\% are derived from the usual independent variations of theoretical $\mu_{F,R}$ scales around $m_{\tata}$ within a factor of two. The PDF uncertainties are negligible compared to those.

\setlength{\tabcolsep}{6.pt}
\begin{table}[htpb!]
\centering
\caption{Ortho-ditauonium cross sections and yields for \pp\ collisions in collider and fixed-target modes at the LHC. The first columns list the NLO cross sections for inclusive p\,p\,$\to \otata+X$ 
and p\,p\,$\to \otata+j$ (with $\pT(\otata)> 2$~GeV) production. The second columns give the expected number of $\otata(\ellell)+j$ events within the LHC detectors acceptances. The last columns list the expected number of $\otata(\ellell)$ events with the additional requirement of a secondary vertex with $L_{xy}> 30~(100)~\mu$m. 
\label{tab:xsec_pp}}
\vspace{0.1cm}
\begin{tabular}{l|cc|cc|cc} \hline
Colliding system,\,$\sqrts$, $\LumiInt$, detector & \multicolumn{2}{c}{$\sigma_\mathrm{NLO}$} & 
\multicolumn{2}{c}{$N(\otata+j)$} & \multicolumn{2}{c}{with $L_{xy}> 30~(100)~\mu$m}\\
& $\otata+X$ & $\otata+j$ & $\otata\to \epem$ & $\otata\to\mumu$ & $\otata\to\epem$ & $\otata\to \mumu$ \\ \hline
\pp\ at 14 TeV, 3 ab$^{-1}$, ATLAS/CMS & $42^{+11}_{-19}$ fb & $18\pm9$ fb & 1100 & 1100 &  130 (10) & 130 (10) \\
\pp\ at 14 TeV, 300 fb$^{-1}$, LHCb &  $42^{+11}_{-19}$ fb & $18\pm9$ fb &  110 &  110 &  5 (--) & 5 (--) \\
\pp\ at 114.6~GeV, 10 fb$^{-1}$, ALICE/LHCb  & $2.2^{+0.3}_{-0.4}$ fb & $1\pm 0.5$ fb & $<$10 & $<$10 & -- & -- \\
\hline
\end{tabular}
\end{table}

\begin{figure}[htpb!]
\centering
\includegraphics[width=0.48\textwidth]{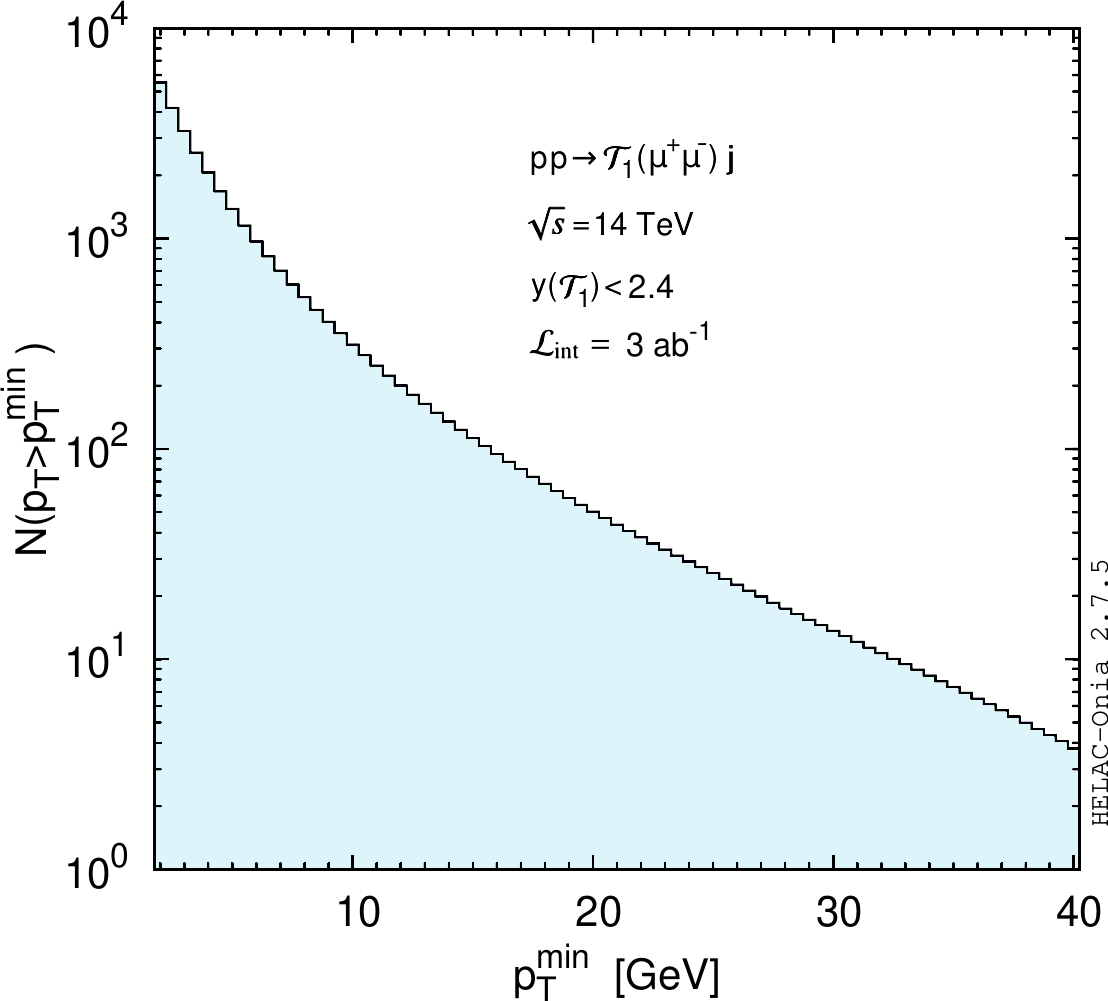}\hspace{0.15cm}
\includegraphics[width=0.48\textwidth]{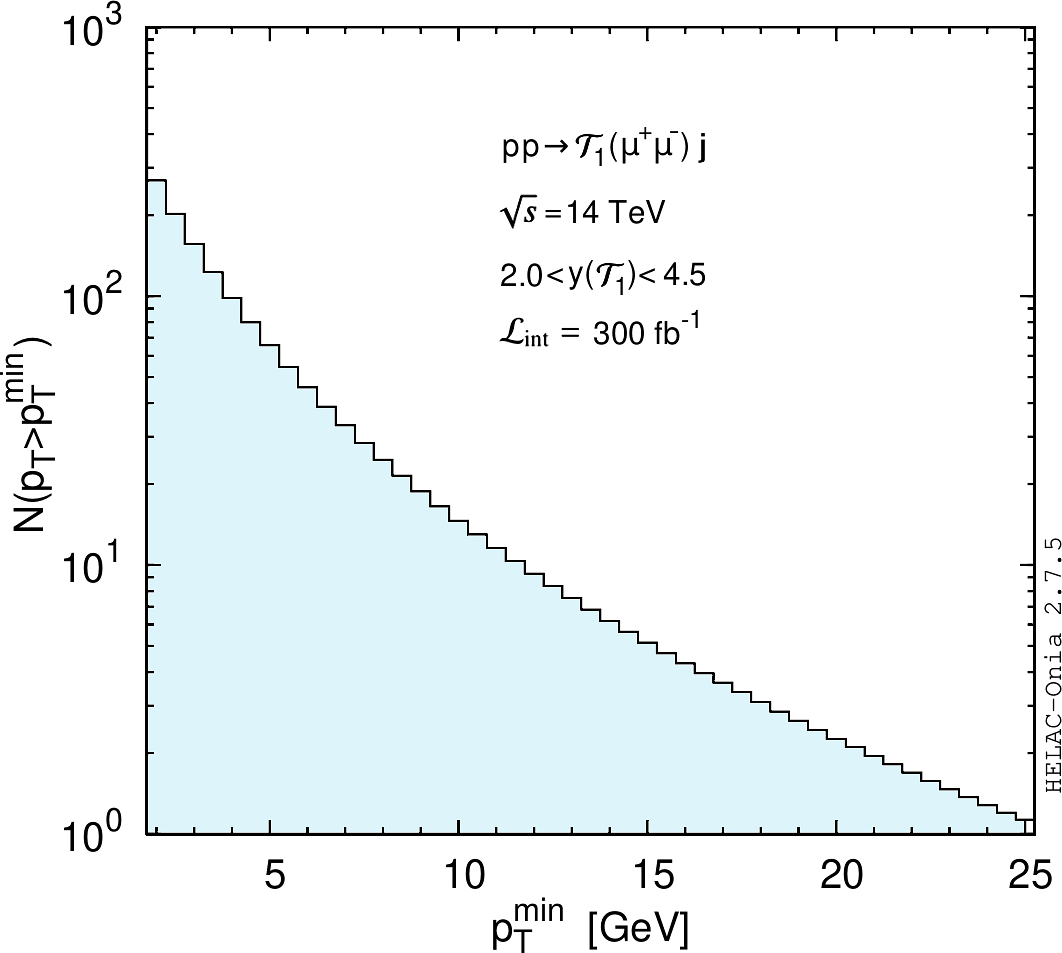}
\caption{Expected number of p\,p\,$\to\otata(\mumu) + j$ events at\,$\sqrts = 14$~TeV above a given minimum $\otata$ $\pT$ within the ATLAS/CMS (left) and LHCb (right) acceptances.
\label{fig:NeventT1jLHC}}
\end{figure}

The process of interest is that where $\otata$ is accompanied by an away-side (mini)jet, p\,p\,$\to \otata(\mumu)+j$, that provides the resonance with a non-zero $\pT$ and allows the use of the displaced $\otata\to\mumu$ decay vertex to identify it above the large Drell--Yan p\,p\,$\to \mumu$ background. 
The second column of Table~\ref{tab:xsec_pp} details the cross section for ditauonium production accompanied by a (mini)jet, $\sigma(\otata+j)$, with $\pT> 2$~GeV. The NLO K-factor for this process is $K=2.3$, as derived also with \madgraph. The produced number of boosted $\otata$ events as function of a given minimum $\pT$ are shown in Fig.~\ref{fig:NeventT1jLHC} for the ATLAS/CMS (left) and LHCb (right) integrated luminosities. For the full HL-LHC programme, one expects about 10 $\otata$ events produced with $\pT \gtrsim 12, 33$~GeV (corresponding to $(\beta\gamma)_{\perp} \approx 3.4, 9.3$ boosts) in LHCb and ATLAS/CMS, respectively.


The expected numbers of actually visible $\otata(\ellell)+j$ events, passing the typical HL-LHC detector acceptance cuts, are listed in the second columns of Table~\ref{tab:xsec_pp}. The requirement of both decay $\ell^\pm$ leptons to be within the ATLAS/CMS (LHCb) acceptance of $|\eta^\ell|<3$ ($2<\eta^\ell<4.5$) and have $\pT^\ell> 2$~GeV ($\pT^\ell> 1$~GeV), reduces the $\otata(\ellell)+j$ yields by about a factor of ten. Thus, a bit more than one thousand events are expected to be produced at ATLAS and CMS, one hundred at LHCb, and just a bunch of them in the fixed-target mode.
The number of expected $\otata$ events with a secondary vertex displaced by at least 30~$\mu$m with respect to the primary vertex, are listed in the last columns of Table~\ref{tab:xsec_pp}. At ATLAS and CMS, about 130~(10) $\otata$ events with $L_{xy}> 30~(100)~\mu$m are expected, indicating that its observation is possible
given that, in principle, there is no known long-lived particle that produces a dimuon peak at $m_{\tata}$. A good control of the reducible combinatorial backgrounds is, however, required. A detailed experimental analysis goes beyond the scope of this paper, but we can outline a few key elements that can facilitate the measurement. Experimental studies of long-lived dimuon resonance searches in ATLAS and CMS~\cite{CMS:2014hka,ATLAS:2018rjc,CMS:2022qej} indicate that displaced dimuons 
originate dominantly from different heavy quarks decays, and partially (mis)reconstructed or misidentified muons from secondary charged pion or kaon decays. The muon isolation requirement strongly suppresses copious backgrounds from dijet and multijet events that yield uncorrelated (either genuine or misidentified) pairs of muons from particles decays within different jets. 
The signed difference in azimuthal angles $\Delta\Phi$ (transverse collinearity angle) between the $\pTmumu$ of the dimuon system and $L_{xy}$ provides an important discriminating variable between signal and backgrounds. 
By requiring $|\Delta\Phi| < \pi/4$ a significant background reduction is achieved, and one is left with contributions from background sources that yield dimuons exclusively or predominantly at small $|\Delta\Phi|$ including: dimuon decays of nonprompt low-mass resonances such as $\jpsi$ or $\psi$(2S) mesons from B-hadron decays; cascade decays of B hadrons; and dimuons formed from a pair of unrelated nonprompt $\mu^\pm$ in the same jet. If well reconstructed, most such background events will not have a $m_{\mumu}$ consistent with the $\tata$ mass and width requirements. Nonetheless, if needed to maximize the analysis sensitivity, multivariate analyses combining multiple discriminating observables in a single powerful discriminator, such as \eg\ a boosted decision tree algorithm, can be further employed.

\subsection{Ditauonium decays into quarkonia-plus-photon, \texorpdfstring{ $\tata_{0,1}\to(\ccbar)_{1,0}+\gamma$}{TT to ccbar+gamma}}

\begin{figure}[htpb!]
\centering
\includegraphics[width=0.95\textwidth]{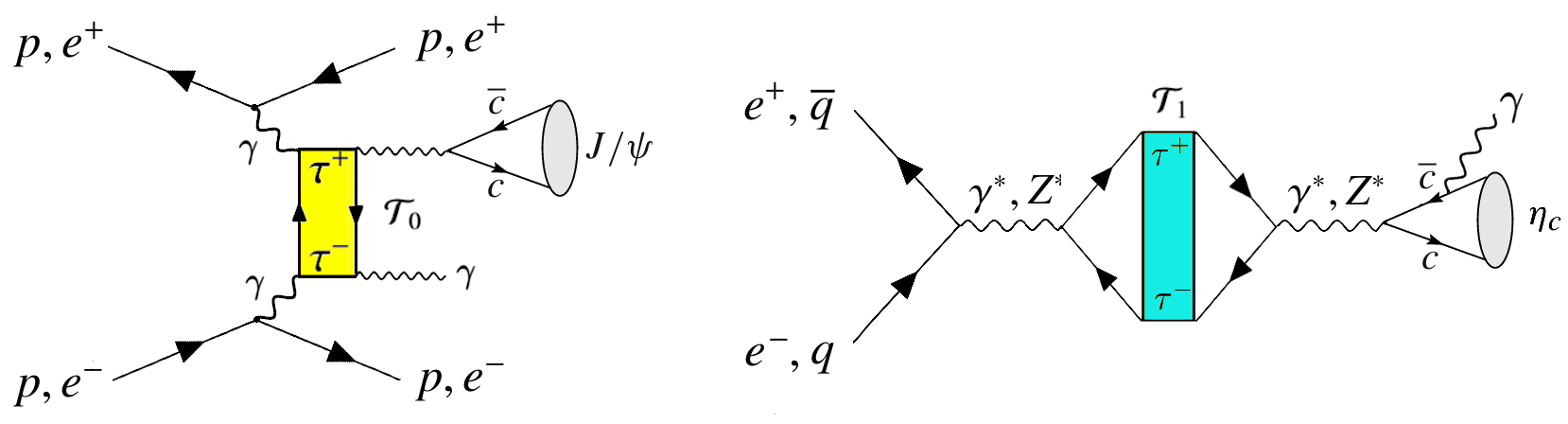}
\caption{Diagrams for para- (left) and ortho- (right) ditauonium production in $\epem$ and \pp\ collisions followed by $(\ccbar)+\gamma$ decays.
\label{fig:ccbargamma_ee_diags}}
\end{figure}

Ditauonium can decay into final states containing a light quark-antiquark pair ($\qqbar=\uubar,\ddbar,\ssbar$), either directly in the ortho-ditauonium case ($\otata\to\qqbar$), or via the Dalitz channel for para-ditauonium where one photon splits into the pair ($\ptata\to\gamma\qqbar$). In principle, the $\ccbar$ decay is not kinematically accessible because the lightest charm D meson has a mass that is more than half the ditauonium mass. It is however possible for ditauonium to decay into light charmonium states plus a photon, via $(\tau^+\tau^-)_0 \to \jpsi+\gamma$ and $(\tau^+\tau^-)_1 \to \eta_c+\gamma$ (Fig.~\ref{fig:ccbargamma_ee_diags}).
The corresponding partial widths can be derived from the NNLO expressions for $\ptata$ Dalitz decays and for $\otata$ diquark-plus-photon decays respectively, multiplied by the standard long-distance matrix elements (LDMEs), $\langle \mathcal{O}^{\jpsi,\eta_c}(^{2s+1}L_{J}^{[c]})\rangle$. LDMEs describe the formation of each quarkonium system in the non-relativistic QCD (NRQCD) framework, as an expansion in terms of the relatively low velocity $v_c$ of the bound charm quarks~\cite{Bodwin:1994jh}. The partial widths for the quarkonium decay modes, suppressed by a $\mathcal{O}(\alpha v_c^3)$ factor and a small phase space compared to the leading decay channels, read
\begin{eqnarray}
\Gamma_{\jpsi\gamma}(\ptata) &=&\frac{4\alpha^6 \pi \, m_\tau}{27\,n^3}\left(1-\frac{m_{\jpsi}^2}{4m_\tau^2}\right) \frac{8\langle O^{\jpsi}(^3S_1^{[1]})\rangle}{m_{\jpsi}^3}\left(1 + \mathcal{O}(v_c^2,\alpha_s,\alpha)\right)\nonumber\\
&=&\frac{3\alpha^4}{n^3}\frac{m_\tau}{m_{\jpsi}}\left(1-\frac{m_{\jpsi}^2}{4m_\tau^2}\right)\Gamma_{\jpsi\to \epem}\left(1 + \mathcal{O}(\alpha)\right)= 6.46\pm 0.12~\mu\text{eV,}
\label{eq:tata_psigamma}\\
\Gamma_{\eta_c\gamma}(\otata) &=&\frac{16\alpha^6 \pi \, m_\tau}{243\,n^3}\left(1-\frac{m_{\eta_c}^2}{4m_\tau^2}\right)\frac{m_{\eta_c}^2}{4m_\tau^2} \frac{8\langle O^{\eta_c}(^1S_0^{[1]})\rangle}{m_{\eta_c}^3}\left(1 + \mathcal{O}(v_c^2,\alpha_s,\alpha)\right)= 1.53_{-0.19}^{+0.89}~\mu\text{eV,} 
\label{eq:tata_etacgamma}
\end{eqnarray}
where the numerical results correspond to the $n=1$ ground states, the QCD coupling evaluated at the Z scale is taken as $\alpha_s(m_\mathrm{Z}) = 0.118$, and $\Gamma_{\jpsi\to \epem}=5.5\pm 0.1$~keV is the $\jpsi$ dielectron partial decay width~\cite{Zyla:2020zbs}. The uncertainty of the $\ptata\to\jpsi\,\gamma$ branching fraction in Eq.~(\ref{eq:tata_psigamma}) is small because it is expressed in terms of the experimentally well-known $\Gamma_{\jpsi\to \epem}$ value. The $\otata\to\eta_c\,\gamma$ decay width, Eq.~(\ref{eq:tata_etacgamma}), uses the $\eta_c$ LDME of $\langle O^{\eta_c}(^1S_0^{[1]})\rangle=0.44$~GeV$^3$ for its numerical evaluation, with an uncertainty estimated by using the different values of the wavefunction at the origin given in~\cite{Eichten:1995ch}.
In addition, relativistic corrections (i.e., higher order in $v_c^2$) and missing higher-order $\alpha_s$ terms can both change this width by $\sim$30\%. The latter two uncertainties have not been included in the estimate quoted here. 
The corresponding decay branching fractions amount to:
\begin{eqnarray}
\mathcal{B}(\ptata\to\jpsi\,\gamma) &=& (0.0271\pm 0.0005)\%,\\
\mathcal{B}(\otata\to\eta_c\,\gamma) &=& \left(0.0048^{+0.0028}_{-0.0006}\right)\%.  
\end{eqnarray}
Such values are so low that experimental searches for ditauonium in the quarkonium$\,+\,$photon decay channels appear hopeless. The $\otata\to \chi_{c}\gamma$ decay is also possible, but it is expected to be further $v_c^4$-suppressed compared to $\otata\to\eta_c\gamma$. In addition, $\otata\to \jpsi+\pi^+\pi^-$ is also kinematically allowed. Its computation requires the knowledge of the $\gamma^*\to \jpsi+\pi^+\pi^-$ transition, and may have larger phase space suppression due to $m_{\tata} \approx m_{\jpsi}+2m_{\pi}$. We refrain ourselves from considering this hadronic channel here.

\section{Summary}

We have evaluated the possibility to experimentally produce and observe ditauonium, the bound state of two tau leptons, in high-energy $\epem$ and hadron collisions. Ditauonium is the heaviest and most compact purely leptonic ``atomic'' system, and remains experimentally unobserved to date. Its experimental study can provide a high-precision extraction of the tau mass, as well as novel QED tests sensitive to physics beyond the standard model, such as \eg\ violations of lepton flavour universality, that may not impact its lighter siblings (positronium and dimuonium). Nine production and decay channels have been studied at four different $\epem$ facilities (BES~III, Belle~II, STCF, and FCC-ee) as well in hadronic and photon-fusion collisions at the LHC. The overall conclusion is that the observation of the para- and/or ortho-ditauonium states is challenging but feasible.\\

Para-ditauonium can be measured at Belle~II and FCC-ee via $\gaga\to\ptata$ photon fusion in its dominant diphoton decay with $\mathcal{B}(\ptata\to\gaga)= 77.72\%$ branching fraction, provided the large background from the overlapping $\chicTwo\to\gaga$ decay can be  precisely controlled in-situ. 
On the other hand, its Dalitz decay with $\mathcal{B}(\ptata\to\ellell\gamma)= 2.31\%$, will likely remain unobserved. 
Ortho-ditauonium can be observed in the $s$-channel in a normal run at a future STCF operating at the $\epem\to\tautau$ threshold. In addition, running with monochromatized beams, leading to a $\mathcal{O}(100$~keV) \cm\ energy spread, the ditauonium cross sections are large, and its mass can be determined from a threshold scan within a precision just given by the \cm\ energy calibration. For a typical $\mathcal{O}(50$~keV) \cm\ energy calibration, the mass of the tau lepton can be obtained with an uncertainty half this value, $\mathcal{O}(25$~keV). Such a ditauonium-based approach will likely become the most precise and accurate experimental means to determine the $\tau$ mass. Ditauonium should thus be incorporated as an integral part of the experimental programme of precision studies of the tau lepton at any future STCF. The ISR production of ortho-ditauonium has too low cross sections at all $\epem$ colliders to be observable over the huge prompt dilepton $\epem\to\ellell(\gamma)$ backgrounds, unless one requires the production of a large-angle photon that boost the resonance decay away from the interaction point. This latter possibility may lead to the observation of a few $\epem\to\otata(\mumu)+\gamma$ events with displaced transverse decays at Belle II.
Finally, the associated production with a jet, p\,p\,$\to \otata(\mumu)+j$, can be observed by ATLAS and CMS at the HL-LHC by measuring its displaced dimuon vertex beyond $L_{xy}\approx 100~\mu$m, with a good control of the large instrumental combinatorial dimuon backgrounds. Last but not least, the very rare decay branching fractions of ditauonium into quarkonium plus~a~photon, $\mathcal{B}(\ptata\to\jpsi\gamma) = 0.0271\%$ 
and $\mathcal{B}(\otata\to\eta_c\gamma) = 0.005\%$, 
have been computed for the first time.\\

\paragraph*{Acknowledgments.---} H.-S.S. thanks S.~Frixione for discussions about the electron structure function. Supports from the European Union's Horizon 2020 research and innovation programme (grant agreements No.\ 824093, STRONG-2020, EU Virtual Access ``NLOAccess'', and No.\ 951754 ``FCC Innovation Study''), the ERC grant (grant agreement ID 101041109, ``BOSON"), the French ANR (grant ANR-20-CE31-0015, ``PrecisOnium''), and the CNRS IEA (grant No.\ 205210, ``GlueGraph"), are acknowledged.




\bibliographystyle{myutphys}
\bibliography{reference}

\end{document}